\documentclass[journal,12pt,letterpaper,draftcls,onecolumn]{IEEEtran}%

\usepackage{amsmath}
\usepackage{amsfonts}
\usepackage{cite}
\usepackage{amssymb}
\usepackage{graphicx}
\usepackage[usenames,dvipsnames]{color}
\usepackage{epstopdf}
\usepackage[all]{xy}
\usepackage[]{mcode}

\usepackage{subcaption}
\usepackage{longtable}

\providecommand{\U}[1]{\protect\rule{.1in}{.1in}}
\pdfoutput=1
\providecommand{\U}[1]{\protect\rule{.1in}{.1in}}

\newtheorem{lemma}{Lemma}
\newtheorem{definition}{Definition}

\newcommand{\qed}{\nobreak \ifvmode \relax \else
      \ifdim\lastskip<1.5em \hskip-\lastskip
      \hskip1.5em plus0em minus0.5em \fi \nobreak
      \vrule height0.75em width0.5em depth0.25em\fi}

\begin{document}

\title{Adaptive OFDM Index Modulation for Two-Hop Relay-Assisted Networks}

\author{Shuping Dang, \IEEEmembership{Student Member, IEEE}, Justin P. Coon, \IEEEmembership{Senior Member, IEEE}, and Gaojie Chen, \IEEEmembership{Member, IEEE} 
  \thanks{
  This work was supported by the SEN grant (EPSRC grant number EP/N002350/1) and the grant from China Scholarship Council  (No. 201508060323).

   The authors are with the Department of Engineering Science, University of Oxford, Parks Road, Oxford, U.K., OX1 3PJ; tel: +44 (0)1865 283 393, (e-mail: \{shuping.dang, justin.coon, gaojie.chen\}@eng.ox.ac.uk).}}

\maketitle

\begin{abstract}
In this paper, we propose an adaptive orthogonal frequency-division multiplexing (OFDM) index modulation (IM) scheme for two-hop relay networks. In contrast to the traditional OFDM IM scheme with a deterministic and fixed mapping scheme, in this proposed adaptive OFDM IM scheme, the mapping schemes between a bit stream and indices of active subcarriers for the first and second hops are adaptively selected by a certain criterion. As a result, the active subcarriers for the same bit stream in the first and second hops can be varied in order to combat slow frequency-selective fading. In this way, the system reliability can be enhanced. Additionally, considering the fact that a relay device is normally a simple node, which may not always be able to perform mapping scheme selection due to limited processing capability, we also propose an alternative adaptive methodology in which the mapping scheme selection is only performed at the source and the relay will simply utilize the selected mapping scheme without changing it. The analyses of average outage probability, network capacity and symbol error rate (SER) are given in closed form for decode-and-forward (DF) relaying networks and are substantiated by numerical results generated by Monte Carlo simulations.
\end{abstract}


\section{Introduction}
\IEEEPARstart{C}{lassic}  modulation schemes, e.g. phase-shift keying (PSK) and quadrature amplitude modulation (QAM), map a sequence of information bits to a modulated data symbol with a certain amplitude and phase, and such a modulated data symbol can be represented by a point in the two-dimensional constellation diagram \cite{proakis}. With rapid increasing demands of reliability and data transmission rate, however, these classic modulation schemes might not be able to satisfy the challenging communication requirements for the next generation networks in terms of spectrum and energy efficiency \cite{6678765}. As a result, there has been a growing interest in design and optimization of modulation schemes which can achieve a high spectrum and/or energy efficiency \cite{6923528}. Two of the most promising schemes are spatial modulation (SM) and orthogonal frequency division multiplexing (OFDM) index modulation (IM), which are capable of adding one extra modulation dimension in addition to the conventional amplitude and phase dimensions \cite{4382913,6094024,6587554,7234862}. SM is designed for multiple-input-multiple-output (MIMO) systems and exploits the spatial dimension, while OFDM IM is mainly applied to OFDM systems by conveying information via the frequency dimension. In practice, OFDM IM is easier to be implemented, since the deployment of MIMO under a limited correlation level among all antennas may not be affordable for most communication systems in terms of cost and physical size \cite{6525429,6823072}, for example  device-to-device (D2D) communication networks in which all nodes are nothing more than idle mobile devices \cite{7577711}. Therefore, OFDM IM is now being regarded as an attractive modulation methodology for the 5G networks \cite{7509396}.

At the meantime, relay-assisted communications have been proved to be an effective approach to provide reliable communication service against propagation attenuation, multipath fading as well as shadowing, and are also regarded as one of the most promising technologies for the next generation cellular networks \cite{7414384}. Moreover, with recent research achievements on multicarrier relay systems, the performance of relay-assisted communications can be further enhanced in terms of multiplexing and/or diversity gains with a moderate system complexity \cite{6473920,4595667,7445895}. Due to the maturity and benefits of multicarrier relay systems, it is of an increasingly high interest to integrate it with other popular communication paradigms \cite{5733966}. In recent years, some works related to SM also considered involving relays to further improve system performance and proposed the concepts of distributed SM (DSM) and dual-hop SM (Dh-SM) \cite{7120187,5956586}.

However, to the best of the authors' knowledge, the link between two-hop relay networks and OFDM IM is still lacking, which motivates us to construct an analytical framework of OFDM IM in two-hop relay-assisted networks in order to integrate the merits of both. Also, motivated by a variety of adaptive modulation and transmit antenna selection schemes proposed for SM and proved to be capable of significantly enhancing system performance \cite{5752793,7055353,6423761}, we would also like to apply the adaptive modulation mechanism in OFDM IM for two-hop relay networks. Based on these considerations, we propose the adaptive OFDM IM for two-hop relay networks in this paper. In particular, a two-hop relay network with one source, one relay and one destination is considered and the mapping scheme for bit sequence to the index of active subcarrier is dynamically selected according to the channel state information (CSI) at source and relay. By such an adaptive mapping scheme selection process, the system reliability can be improved. 

In spite of mapping scheme adaptation, our proposed scheme is different from those traditional OFDM IM schemes, in which only a fixed number of subcarriers are active in each instant and thus will result in a low spectral efficiency \cite{7763523,6162549,6587554}. Instead, an efficient activation mechanism with multiple active subcarriers is used in our proposed scheme to map the incoming bit stream to a corresponding subcarrier activation pattern. On the other hand, despite the higher spectral efficiency provided by adopting such an efficient activation mechanism with an indeterministic number of active subcarriers, there is a small possibility that all subcarriers are inactive. In this special case, the amplitude and phase modulation (APM) data symbol cannot be transmitted. We term this \textit{zero-active subcarrier dilemma}, which hinders the utilization of a variable number of active subcarriers. In this paper, an additional benefit brought by the subcarrier adaptation is that we can employ a dual-mode transmission protocol to solve the zero-active subcarrier dilemma.

Following by these core ideas, the contributions of this paper are listed infra:
\begin{enumerate}
\item We propose an adaptive OFDM IM scheme for two-hop relay networks, by which the mapping schemes are dynamically selected based on instantaneous CSI and a mechanism allowing reducing throughput for gaining reliability can be constructed. Meanwhile, for implementation in practice, we propose two feasible mapping scheme selection methodologies, which can be applied depending on the design specification of the relay node, i.e. whether the relay is capable of performing mapping scheme selection or not.
\item We propose a dual-mode transmission protocol which can solve the zero-active subcarrier dilemma and transmit the APM symbol when all selected subcarriers are inactive.
\item We analyze the outage performance of the proposed modulation schemes and derive the exact expressions and asymptotic expressions in the high signal-to-noise ratio (SNR) region for the average outage probability in closed form; we analyze the average network capacity and derive its closed-form expression; we also investigate the error performance and give a closed-form approximation of the average symbol error rate (SER).
\end{enumerate}

The rest of this paper is organized as follows. In Section \ref{sm}, the system model of adaptive OFDM IM in two-hop networks is introduced. After that, we specify the mapping scheme selection and performance evaluation metrics in Section \ref{mappingschemeselection}. The analyses of average outage performance, network capacity and error performance of the proposed systems are carried out in Section \ref{opa}, Section \ref{anca} and Section \ref{epa}, respectively. Then, all analyses are numerically verified and the simulation results are discussed in Section \ref{nr}. Finally, Section \ref{c} concludes the paper.

\section{System Model}\label{sm}

\begin{figure}[!t]
\centering
\includegraphics[width=6.0in]{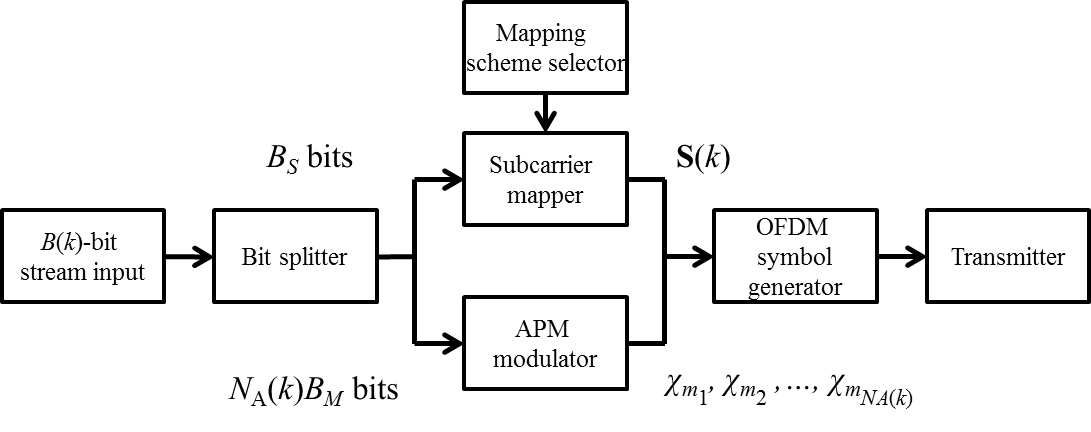}
\caption{System block diagram of the source (similar to the transmission part of the relay).}
\label{sys1}
\end{figure}

\subsection{System framework}\label{sfsmdsa232356565}
Consider a two-hop OFDM IM system with one source, one relay, one destination and $N_T$ subcarriers, which is operating in slow frequency-selective Rayleigh fading channels. Also, $N_S$ out of the $N_T$ subcarriers will be selected by a certain criterion to form a \textit{mapping scheme} (a.k.a. a \textit{codebook}) for OFDM IM, where $1 \leq N_S< N_T$. In each instant, there is a variable-length equiprobable stream $\mathbf{b}_S(k)+\sum_{t=1}^{N_A(k)}\mathbf{b}_M(m_t)$ required to be transmitted at the source, where $m_t\in\mathcal{M}=\{1,2,\dots,M\}$ denoting the index of the chosen APM constellation symbol for the $t$th active subcarrier and $M$ is the order of APM; $k\in\mathcal{K}=\{1,2,3,\dots,2^{N_S}\}$ denoting the index of the subcarrier activation pattern and $N_A(k)$ is the corresponding number of active subcarriers. Supposing that the lengths of $\mathbf{b}_S(k)$ and $\mathbf{b}_M(m_t)$ are $B_S$ and $B_M$, the length of the entire variable-length stream is $B(k)=B_S+N_A(k)B_M$. The $B_S$-bit stream $\mathbf{b}_S(k)$ is modulated by the subcarrier activation pattern, and $N_A(k)$ out of $N_S$ subcarriers will be activated to convey the APM symbols generated according to these $B_M$-bit streams $\mathbf{b}_M(m_1)$, $\mathbf{b}_M(m_2)$,$\dots$,$\mathbf{b}_M(m_{N_A(k)})$ by the conventional $M$-ary APM scheme (e.g. $M$-PSK or $M$-QAM). Note that, normally the mapping scheme is preset in classic OFDM IM systems and remains fixed, but in this paper, we assume it can be dynamically selected according to the CSI at both source and relay. More details of the mapping scheme and its adaptation mechanisms will be given latter in Section \ref{mappingschemeselection} and Fig. \ref{shrinking}. 

Meanwhile, we assume that a single bit in $\mathbf{b}_S(k)$ represents the activation state of a given subcarrier in an on-off keying (OOK) manner. In other words, for a given $B_S$-bit stream $\mathbf{b}_S(k)$, the selected subcarriers having the same sequence numbers with bits `1' will be activated, on which $N_A(k)$ different APM symbols are transmitted. That is to say, $N_S=B_S$ and the number of active subcarriers $N_A(k)$ is equal to the \textit{hamming weight} of the $B_S$-bit stream $\mathbf{b}_S(k)$. More specifically, the activation state matrix can be designed as
\begin{equation}\label{activationstatematrix}\small
\mathbf{S}(k)=\mathrm{diag}\{b_S(k,1),b_S(k,2),\dots,b_S(k,N_S)\},
\end{equation}
where $b_S(k,n)$ is either `0' or `1' denoting the $n$th entry in $\mathbf{b}_S(k)$, $\forall~n\in\{1,2,\dots,N_S\}$. As a result, by OFDM IM, the information contained in the entire $B(k)$-bit stream is now conveyed by both of $N_A(k)$ $M$-ary constellation symbols $\{\chi_{m_t}\}$ and the activation state matrix $\mathbf{S}(k)$.

Here we further assume that there is no direct transmission link between source and destination due to deep fading or severe propagation attenuation and a half-duplex forwarding protocol is adopted at the relay, so that two orthogonal temporal phases are required for a complete symbol transmission from source to destination. In the first phase, the generated OFDM IM symbol at the source is transmitted to the relay in slow frequency-selective Rayleigh fading channels. Assuming decode-and-forward (DF) relaying protocol is adopted at the relay, once the distorted symbol is received by the relay, it will be first decoded and regenerated according to the relay's own mapping scheme for the second hop. Then the relay transmits the regenerated symbol to the destination. Finally, the distorted version of the new symbol is received by the destination and then decoded to retrieve the complete $B(k)$-bit stream. For clarity, the system block diagram of the source is illustrated in Fig. \ref{sys1}, which is also highly similar to the transmission part of the relay.

\subsection{Channel model}
We assume the wireless channels in the first and second hops to be slow frequency-selective faded and their gains to be exponentially distributed but with different means $\mu_1$ and $\mu_2$ due to different propagation attenuations. Note that, the slow or quasi-static property referred to here indicates that the channel gains are random but would remain unchanged for a sufficiently large period of time  \cite{tse2005fundamentals}. Based on such an assumption, the overheads caused by transmitting the selected mapping scheme via feedforward links to relay and destination for decoding purposes are negligible \cite{5752793}. This justifies the feasibility of our proposed adaptive OFDM IM scheme in practice. Denoting the set of all subcarriers as $\mathcal{N}=\{1,2,\dots,N_T\}$, the probability density function (PDF) and the cumulative distribution function (CDF) of the channel gain $|h_i(n)|^2$, $\forall~n\in\mathcal{N}$, can be written as
\begin{equation}\label{channelpdfcdf}\small
f_{i}(s)=\mathrm{exp}\left(-s/\mu_i\right)/\mu_i~~\Leftrightarrow~~F_{i}(s)=1-\mathrm{exp}\left(-s/\mu_i\right)
\end{equation}
where $i\in\{1,2\}$ denotes the index of the first or second hop. Meanwhile, because of the fundamental assumptions of OFDM, it is supposed that these $N_T$ channel gains in each hop are statistically independent\footnote{We assume a block-fading model in frequency akin to systems that employ a resource block frame/packet structure (e.g., Long-Term Evolution (LTE)), and hence the independent and identically distributed (i.i.d.) assumption in frequency for each hop can be justified \cite{7445895}.}.

\subsection{APM scheme}
In this paper, we adopt $M$-PSK as the APM scheme\footnote{As has been proved for SM systems and similarly for OFDM IM systems, a constant-envelope APM is preferable over non-constant-envelope APM when the spatial or frequency indices are employed for information transmission, since it can offer a higher power efficiency and a reduced-complexity decoding process \cite{6142142,7330022}. In other words, $M$-PSK outperforms $M$-QAM. This is the reason of using $M$-PSK in this paper.}. Meanwhile, the analytical results in the following sections can also be used for $M$-QAM cases if some precoding schemes (e.g. dynamic power allocation on constellation points) can be applied on each data symbol so that a constant envelope is achieved \cite{luna2013constellation,gonzalez2012pre}. Without loss of generality, we normalize the data symbol by $\chi_{m_t}\chi_{m_t}^*=1$, $\forall~m_t\in\mathcal{M}$.

\section{Mapping Scheme Selections and Performance Evaluation Metrics}\label{mappingschemeselection}
Now let us focus on the key property of the proposed system, i.e. dynamical mapping scheme selection according to CSI. First, here we assume CSI can be perfectly known at source and relay without delay and overhead. Then, the set of all mapping schemes is denoted by $\mathcal{C}$ and the cardinality of $\mathcal{C}$, i.e. the total number of possible mapping schemes is given by $\mathrm{Card}(\mathcal{C})=\binom{N_T}{N_S}$, where $\binom{\cdot}{\cdot}$ denotes the binomial coefficient. It should be noted that the OFDM blocks are normally generated by inverse fast Fourier transform (IFFT), which will require $N_T=2^\zeta$, $\zeta\in\mathbb{N}^+=\{1,2,\dots\}$.

\subsection{Signal model and dual-mode transmission protocol}
In order to perform mapping scheme selection, we have to construct the signal model and transmission protocol as well as some important performance metrics. Now, let us first focus on the construction of the transmit OFDM block (a.k.a. transmit signal vector) before adopting a certain mapping scheme. Assuming a cyclic prefix (CP) with sufficient length is applied to the time-domain transmit OFDM block produced by a $N_T$-point IFFT, the OFDM signaling model can be considered at a subcarrier level \cite{7247508}. A general form of the $N_T \times 1$ transmit OFDM block in frequency domain without interblock interference can be written as
\begin{equation}\small
\mathbf{x}=[x(1),x(2),\dots,x(N_T)]^T\in\mathbb{C}^{N_T\times 1},
\end{equation}
where $(\cdot)^T$ denotes the matrix transpose operation.

When selecting an arbitrary $c$th mapping scheme, $c\in\mathcal{C}$, which selects $N_S$ out of $N_T$ subcarriers to form a subset $\mathcal{N}_S(c)\subset\mathcal{N}$ for OFDM IM, the reduced OFDM block determined by the $B(k)$-bit stream is given by
\begin{equation}\label{545msmdx2}\small
\mathbf{x}(k)=[x(m_1,1),x(m_2,2),\dots,x(m_{N_S},N_S)]^T\in\mathbb{C}^{N_S\times 1},
\end{equation}
where
\begin{equation}\small
x(m_n,n)=\begin{cases}
\chi_{m_n}, ~~~~n\in\mathcal{N}_A(k)\\
0,~~~~~~~~\mathrm{otherwise}
\end{cases}
\end{equation}
corresponds to the data symbol transmitted on the $n$th subcarrier, and $\mathcal{N}_A(k)\subseteq\mathcal{N}_S(c)$ is the subset of $N_A(k)$ active subcarriers out of $N_S$ selected subcarriers when $\mathbf{b}_S(k)$ is transmitted.

Now, in order to avoid the \textit{zero-active subcarrier dilemma}, which is defined as the case when all selected subcarriers are required to be inactive (i.e. $\mathbf{b}_S(1)=\{0,0,\dots,0\}$ and thus $N_A(1)=0$), we propose a \textit{dual-mode transmission protocol}, which consists of the regular transmission mode and the complementary transmission mode.

If $\mathbf{S}(k)\neq \mathbf{0}_{N_S\times N_S}$, i.e. $\mathbf{b}_S(k)$ is not an all-zero bit stream, consequently, the regular transmission mode is activated and the received signal for the $i$th hop can be expressed by\cite{7469311}
\begin{equation}\label{dsa45232312}\small
\begin{split}
\mathbf{y}_i(k)&=[y_i(m_1,1),y_i(m_2,2),\dots,y_i(m_{N_S},N_S)]^T\\
&=\sqrt{\frac{P_t}{N_A(k)}}\mathbf{H}_i(c)\mathbf{x}(k)+\mathbf{w}_i\in\mathbb{C}^{N_S\times 1},
\end{split}
\end{equation}
where $\mathbf{w}_i=[w_i(1),w_i(2),\dots,w_i(N_S)]^T\in\mathbb{C}^{N_S\times 1}$ is the vector of independent complex additive white Gaussian noise (AWGN) samples on each subcarrier at the $i$th hop, whose entries obey $\mathcal{CN}(0,N_0)$, and $N_0$ is the noise power; $\mathbf{H}_i(c)=\mathrm{diag}\{h_i(c,1),h_i(c,2),\dots,h_i(c,N_S)\}\in\mathbb{C}^{N_S\times N_S}$ is a $N_S\times N_S$ diagonal channel state matrix for the $i$th hop when the $c$th mapping scheme is selected; $P_t$ is a uniform transmit power adopted at both source and relay. Due to the normalization of APM symbols, we obtain the received SNR in the $i$th hop for the $n$th subcarrier as
\begin{equation}\label{receivedsnr}\small
\gamma_i(k,n)=\frac{P_t}{N_A(k)N_0}b_S(k,n)|h_i(c,n)|^2,~~~~~\forall~n\in\mathcal{N}_S(c).
\end{equation}

However, for $\mathbf{S}(1)= \mathbf{0}_{N_S\times N_S}$ (we denote $k=1$ for this case), i.e. $\mathbf{b}_S(1)$ is an all-zero bit stream,  one of those $N_T-N_S$ unselected subcarriers $\tilde{n}_i$ will all be activated to undertake the transmission of data symbol $\chi_{m_{\tilde{n}}}$ in the $i$th hop. In this case, the received signal transmitted on the complementary subcarrier in the $i$th hop becomes
\begin{equation}\small
y_i(1,\tilde{n}_i)=h_i(\tilde{n}_i)\chi_{m_{\tilde{n}}}+w_i(\tilde{n}_i),
\end{equation}
where $w_i(\tilde{n}_i)$ is a complex AWGN on the complementary subcarrier, and the SNRs of the complementary subcarrier as well as other selected subcarriers become
\begin{equation}\small
\gamma_i(1,\tilde{n}_i)=\frac{P_t}{N_0}|h_i(\tilde{n}_i)|^2~~\mathrm{and}~~\gamma_i(1,n)=0,~\forall~n\in\mathcal{N}_S(c).
\end{equation}
We term this special case \textit{complementary transmission} mode, so that at least one data symbol $\chi_{m_{\tilde{n}}}$ can still be transmitted on the standby subcarrier $\tilde{n}_i$ when all selected subcarriers are inactive\footnote{The information conveyed by $\chi_{m_{\tilde{n}}}$ can be crucial for system coordination, e.g. synchronization.}. Based on the specification above for the dual-mode transmission protocol, we can have the average transmission rate in bit per channel use (bpcu) as follows
\begin{equation}\label{56422s}\small
\begin{split}
\bar B&=\log_2\left(2^{N_S}\right)+\underset{k\in\mathcal{K}}{\mathbb{E}}\left\lbrace \max\left\lbrace 1,N_A(k)\right\rbrace \log_2(M)\right\rbrace=N_S+\frac{\log_2(M)}{2^{N_S}}\left(1+2^{N_S-1}N_S\right)~~[\mathrm{bpcu}],\\
\end{split}
\end{equation}
where $\mathbb{E}\{\cdot\}$ denotes the expected value of the random variable enclosed.

\subsection{Decentralized mapping scheme selection methodology}
Here, we propose two feasible mapping scheme selection methodologies, which can be applied depending on the design specification of the relay node, i.e. whether the relay is capable of performing mapping scheme selection or not. We first introduce the decentralized mapping scheme as follows. When both source and relay can get access to perfect CSI and carry out the subcarrier adaptation process, in order to optimize the outage performance and error performance, the mapping scheme selections are independently performed at the source and relay according to the criterion\footnote{We denote the index of all-one $B_S$-bit stream as $k=2^{N_S}$, i.e. $\mathbf{b}_S(2^{N_S})=[1,1,\dots,1]$. In this case, all $N_S$ selected subcarriers will be active, so that a comprehensive mapping scheme selection considering the qualities of all channels can be performed.}
\begin{equation}\label{mappselect1}\small
\hat{c}_i=\underset{c\in\mathcal{C}}{\arg\max}\left\lbrace\sum_{n\in\mathcal{N}_S(c)}\gamma_i(2^{N_S},n)\right\rbrace,
\end{equation}
where $\hat{c}_i$ denotes the index of the optimal mapping scheme selected for the $i$th hop. Because the mapping scheme selections are independently performed at both source and relay, we term this selection methodology the \textit{decentralized mapping scheme selection}.

Subsequently, for the $i$th hop, the selection criterion of the complementary subcarrier is given by
\begin{equation}\label{xuanzejizhibuchong}\small
\tilde{n}_i=\underset{n\in\mathcal{N}\setminus\mathcal{N}_S(\hat{c}_i)}{\arg\max}|h_i(n)|^2.
\end{equation}

\subsection{Centralized mapping scheme selection methodology}

However, the decentralized mapping scheme selection methodology will result in a high system complexity at the relay node, which might not be practical as the relay is normally an idle user mobile device with relatively limited processing capability \cite{6211487}. We thereby propose an alternative suboptimal mapping scheme selection methodology which is performed only at the source and thus relaxes the design requirements of the relay. Because the mapping scheme selection is only performed at the source, we term this the \textit{centralized mapping scheme selection}. Specifically, the mapping scheme is only selected at the source and will be simply followed by the relay. The mapping scheme selection criterion is given by
\begin{equation}\label{mappselect3}\small
\hat{c}=\underset{c\in\mathcal{C}}{\arg\max}\left\lbrace\sum_{n\in\mathcal{N}_S(c)}\min\left\lbrace\gamma_1(2^{N_S},n),\gamma_2(2^{N_S},n)\right\rbrace\right\rbrace.
\end{equation}
In a similar manner to the decentralized case, we can select the contemporary subcarrier for a whole link by the criterion
\begin{equation}\small
\tilde{n}=\underset{n\in\mathcal{N}\setminus\mathcal{N}_S(\hat{c})}{\arg\max}\min\left\lbrace|h_1(n)|^2,|h_2(n)|^2\right\rbrace.
\end{equation}

\begin{figure}[!t]
\centering
\includegraphics[width=3.0in]{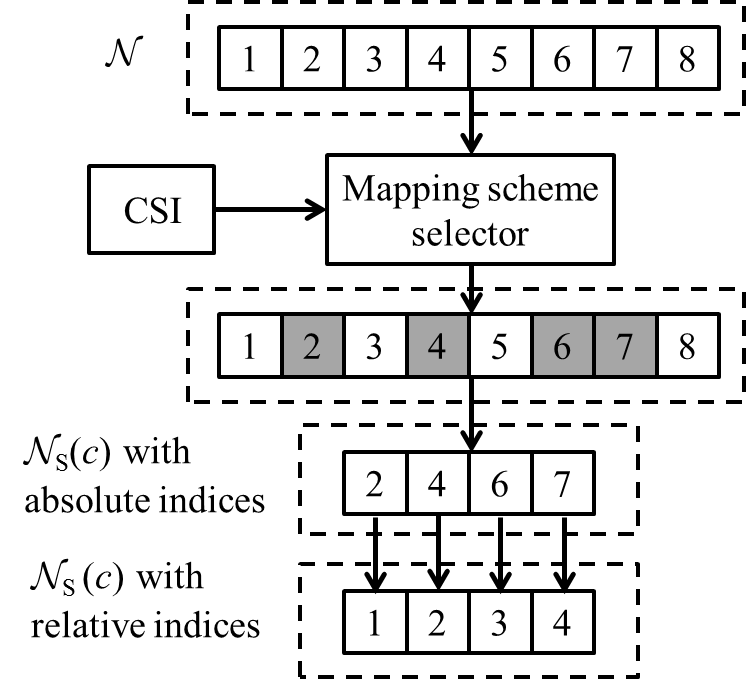}
\caption{An example of mapping scheme selection process from $\mathcal{N}$ to $\mathcal{N}_S(c)$, when $N_T=8$ and $N_S=4$.}
\label{shrinking}
\end{figure}

By either the decentralized or centralized mapping scheme selection methodology, as long as the $c$th mapping scheme is selected, the $B_S$-bit stream $\mathbf{b}_S(k)$ will be mapped to activate a subset of $\mathcal{N}_A(k)$ from $\mathcal{N}_S(c)$ subcarriers. To achieve this, we first need to relabel the subcarriers in $\mathcal{N}_S(c)$ by the new indices in ascending order. To be clear, the indices in the original full set $\mathcal{N}$ are termed \textit{absolute indices}, whereas the indices in the selected subset $\mathcal{N}_S(c)$ are termed \textit{relative indices}\footnote{Because the values of indices \textit{per se} do not affect the mathematical analysis, with a slight abuse of notation, we do not specifically distinguish them and denote a generic subcarrier index as $n$ in this paper.}. An example of the mapping scheme selection process from $\mathcal{N}$ to $\mathcal{N}_S(c)$ can be observed in Fig. \ref{shrinking}, given $N_T=8$ and $N_S=4$. Furthermore, for clarity, an example of the OFDM IM mapping table when $B_M=1$ and $B_S=2$ is given in Table \ref{exmtable} (Because $B_M=1$, BPSK is adopted for APM and here we stipulate that data symbols $+1$ and $-1$ correspond to bit `1' and `0', respectively).

\begin{table}[!t]
\renewcommand{\arraystretch}{1.3}
\caption{An example of the OFDM IM mapping table when $B_M=1$ and $B_S=N_S=2$.}
\label{exmtable}
\centering
\begin{tabular}{c|c|c|c|c|c}
$\mathbf{b}_S(k)$&$\mathcal{N}_A(k)$&$\sum_{t=1}^{N_A(k)}\mathbf{b}_M(m_t)$&  $\mathbf{S}(k)$& $\mathbf{x}(k)$ & $\chi_{m_{\tilde{n}}}$\\
\hline
$[0,0]$&$\varnothing$&$[0]$& $\mathrm{diag}\{0,0\}$ &$[0,0]^T$* & -1\\
$[0,0]$&$\varnothing$&$[1]$&  $\mathrm{diag}\{0,0\}$ &$[0,0]^T$*& +1\\
$[0,1]$&$\{1\}$&$[0]$&  $\mathrm{diag}\{0,1\}$ &$[0,-1]^T$& 0\\
$[0,1]$&$\{1\}$&$[1]$& $\mathrm{diag}\{0,1\}$ &$[0,+1]^T$& 0\\
$[1,0]$&$\{2\}$&$[0]$&  $\mathrm{diag}\{1,0\}$ &$[-1,0]^T$& 0\\
$[1,0]$&$\{2\}$&$[1]$&  $\mathrm{diag}\{1,0\}$ &$[+1,0]^T$& 0\\
$[1,1]$&$\{1,2\}$&$[0,0]$&  $\mathrm{diag}\{1,1\}$ &$[-1,-1]^T$& 0\\
$[1,1]$&$\{1,2\}$&$[0,1]$&  $\mathrm{diag}\{1,1\}$ &$[-1,+1]^T$& 0\\
$[1,1]$&$\{1,2\}$&$[1,0]$& $\mathrm{diag}\{1,1\}$ &$[+1,-1]^T$& 0\\
$[1,1]$&$\{1,2\}$&$[1,1]$&  $\mathrm{diag}\{1,1\}$ &$[+1,+1]^T$& 0\\
\end{tabular}
\\ \ \\ \ \\ * The complementary transmission mode will be activated in this case.
\end{table}

\subsection{Performance evaluation metrics}\label{pem}
\subsubsection{Average outage probability}
We define the outage event of the proposed systems as follows \cite{6157252}.
\begin{definition}\label{defoutages323x}
An outage occurs when the SNR of any of the active subcarriers in either the first or the second hop falls below a preset outage threshold $s$.
\end{definition}

By fundamental probability theory, we can determine the conditional outage probability when $\mathbf{b}_S(k)$ is transmitted by
\begin{equation}\label{212outasdsa1e23}\small
P_o(s|k)=\begin{cases}
\mathbb{P}\left\lbrace \underset{i=1,2}{\bigcup}\left\lbrace\gamma_i(1,\tilde{n}_i)<s\right\rbrace\right\rbrace,~~~~~~~~~~~~~~~\mathrm{if}~k=1\\
\mathbb{P}\left\lbrace \underset{i=1,2}{\bigcup}\left\lbrace \underset{n\in\mathcal{N}_A(k)}{\bigcup}\left\lbrace\gamma_i(k,n)<s\right\rbrace\right\rbrace\right\rbrace,~\mathrm{if}~k>1
\end{cases}
\end{equation}
where $\mathbb{P}\{\cdot\}$ denotes the probability of the event enclosed. Therefore, the \textit{average outage probability} over $k$ can be written as
\begin{equation}\label{2d1sa23142}\small
\bar P_o(s)=\underset{k\in\mathcal{K}}{\mathbb{E}}\left\lbrace P_o(s|k)\right\rbrace,
\end{equation}
which is an important metric used in this paper to characterize the system reliability.

\subsubsection{Average network capacity}
To jointly consider throughput and reliability, we should also take network capacity into consideration to further investigate the performance of the proposed systems. According to the max-flow min-cut theorem \cite{5071270}, the \textit{average network capacity} in two-hop networks is given by
\begin{equation}\small
\begin{split}
&\bar{C}=\underset{\begin{subarray}{c}\mathbf{H}_i(\hat{c}_i),h_i(\tilde{n}_i)\\i\in\{1,2\},k\in\mathcal{K}\end{subarray}}{\mathbb{E}}\left\lbrace C(k)\right\rbrace~~[\mathrm{bit/s/Hz}],
\end{split}
\end{equation}
where $C(k)$ is the network capacity when $\mathbf{b}_S(k)$ is transmitted and can be written as
\begin{equation}\label{dsadoutsctutasn}\small
C(k)=\begin{cases}
\frac{1}{2}\min\left\lbrace\log_2\left(1+\gamma_1(1,\tilde{n}_1)\right),\log_2\left(1+\gamma_2(1,\tilde{n}_2)\right)\right\rbrace,~~~~~~~~~~~~~~~~~\mathrm{if}~k=1\\
\underset{n\in\mathcal{N}_A(k)}{\sum}\frac{1}{2}\min\left\lbrace\log_2\left(1+\gamma_1(k,{n})\right),\log_2\left(1+\gamma_2(k,{n})\right)\right\rbrace,~~~~~~~~~~\mathrm{if}~k>1\\
\end{cases}
\end{equation}

\subsubsection{Average symbol error rate}
To analyze the error performance, we have to first specify the detection method used in our proposed systems. Although the complexity of the ML detection will increase exponentially with the APM order $M$ and the number of selected subcarriers $N_S$, it can provide the optimal error performance and serves as a good reference to other suboptimal detection methods \cite{4601434}. Therefore, we adopt the ML detection method at the relay and destination to demodulate the received signal and retrieve the transmitted bits. However, due to the dual-mode transmission protocol adopted in this paper, the ML detection method has to be tailored accordingly. In order to effectively perform detection, we first need to construct a concatenated symbol block by $\mathbf{x}(k)$ and $\chi_{m_{\tilde{n}}}$ as $\mathbf{X}(k)=\langle\mathbf{x}(k),\chi_{m_{\tilde{n}}}\rangle$, where $\langle\cdot,\cdot\rangle$ represents the vector/diagonal matrix concatenation operation\footnote{Here we define: for arbitrary vectors $\mathbf{u}_1=[u_1(1),u_1(2),\dots,u_1(n)]^T$ and $\mathbf{u}_2=[u_2(1),u_2(2),\dots,u_2(m)]^T$, $\langle\mathbf{u}_1,\mathbf{u}_2\rangle=[u_1(1),u_1(2),\dots,u_1(n),u_2(1),u_2(2),\dots,u_2(m)]^T$; for arbitrary diagonal  matrices $\mathbf{U}_1=\mathrm{diag}\{u_1(1),u_1(2),\dots,u_1(n)\}$ and $\mathbf{U}_2=\mathrm{diag}\{u_2(1),u_2(2),\dots,u_2(m)\}$, $\langle\mathbf{U}_1,\mathbf{U}_2\rangle=\mathrm{diag}\{u_1(1),u_1(2),\dots,u_1(n),u_2(1),u_2(2),\dots,u_2(m)\}$.}. The set of all possible $\mathbf{X}(k)$ is denoted as $\mathcal{X}$ and $\mathrm{Card}(\mathcal{X})=M+\sum_{n=1}^{N_S}\binom{N_S}{n}M^n$. Subsequently, the detection criterion of the ML detector for the received concatenated signal $\dot{\mathbf{Y}}_i(\dot{k})=\langle\dot{\mathbf{y}}_i(\dot{k}),y_i(1,\tilde{n}_i)\rangle$ distorted from the authentic transmitted concatenated OFDM IM symbol $\dot{\mathbf{X}}(\dot{k})$ in the $i$th hop is given by
\begin{equation}\small
\begin{split}
&\hat{\mathbf{X}}(\hat{k})=\underset{{\mathbf{X}}({k})}{\arg\min}\begin{Vmatrix}\dot{\mathbf{Y}}_i(\dot{k})-\sqrt{\frac{P_t}{N_A(k)}}\langle\mathbf{H}_i(\hat{c}_i),h_i(\tilde{n}_i)\rangle\mathbf{X}(k)\end{Vmatrix}_F,
\end{split},~~~~~~~~~~
\end{equation}
where $\begin{Vmatrix}\cdot\end{Vmatrix}_F$ denotes the Frobenius norm of the enclosed matrix/vector.

Hence, the conditional SER for the $i$th hop can be written as
\begin{equation}\label{shijiugongset1}\small
\begin{split}
&P_{e-i}(\dot{\mathbf{X}}(\dot{k})\vert \mathbf{H}_i(\hat{c}_i),h_i(\tilde{n}_i))=\underset{\hat{\mathbf{X}}(\hat{k})\neq \dot{\mathbf{X}}(\dot{k})}{\sum} P_{e-i}(\dot{\mathbf{X}}(\dot{k})\rightarrow \hat{\mathbf{X}}(\hat{k})\vert \mathbf{H}_i(\hat{c}_i),h_i(\tilde{n}_i)).
\end{split}
\end{equation}
Subsequently, the unconditional SER can be expressed as
\begin{equation}\label{shijiugongset2}\small
P_{e-i}(\dot{\mathbf{X}}(\dot{k}))=\underset{\mathbf{H}_i(\hat{c}_i),h_i(\tilde{n}_i)}{\mathbb{E}}\left\lbrace P_{e-i}(\dot{\mathbf{X}}(\dot{k})\vert \mathbf{H}_i(\hat{c}_i),h_i(\tilde{n}_i))\right\rbrace.
\end{equation}
Due to the bottleneck effect of a two-hop network, the end-to-end SER can be approximated by\footnote{For simplicity, we neglect the circumstance that an erroneously estimated and retransmitted signal at the relay can be again `erroneously' estimated to the correct signal at the destination, as the probability is small.}
\begin{equation}\label{shijiugongset3}\small
P_{e}(\dot{\mathbf{X}}(\dot{k}))\approx P_{e-1}(\dot{\mathbf{X}}(\dot{k}))+P_{e-2}(\dot{\mathbf{X}}(\dot{k}))-P_{e-1}(\dot{\mathbf{X}}(\dot{k}))P_{e-2}(\dot{\mathbf{X}}(\dot{k})).
\end{equation}
Now, we can define the average SER as
\begin{equation}\label{SERzuizhongban}\small
\bar P_{e}=\underset{\dot{\mathbf{X}}(\dot{k})\in\mathcal{X}}{\mathbb{E}}\left\lbrace P_{e}(\dot{\mathbf{X}}(\dot{k}))\right\rbrace,
\end{equation}
which will be used in this paper to evaluate the error performance of the proposed systems.

\subsection{Comparison benchmarks}
\subsubsection{OFDM IM without mapping scheme adaptation}
Without mapping scheme adaptation, the OFDM IM scheme will activate $N_T/2$ out of $N_T$ subcarriers in each instant to convey information \cite{6587554}.  The transmission rate of the OFDM IM without mapping scheme adaptation is given by
\begin{equation}\label{465341653452132}\small
B_{classic}=\frac{N_T}{2}\log_2(M)+\lfloor\log_2\binom{N_T}{{N_T}/{2}}\rfloor~~~~~~[\mathrm{bpcu}],
\end{equation}
where $\lfloor\cdot\rfloor$ represents the floor function.

It can be seen by comparing (\ref{56422s}) with (\ref{465341653452132}) that the transmission rate of the OFDM IM scheme without mapping scheme adaptation is higher than that with mapping scheme adaptation. Another merit of the OFDM IM without mapping scheme adaptation is that it always activates $N_T/2$ subcarriers, so that the zero-active subcarrier dilemma does not exist. However, because the OFDM IM scheme without mapping scheme adaptation utilizes the index of subcarrier set to convey information, it is expected that some possible subcarrier sets will never be used in this scheme and therefore the spectral efficiency is degraded without any benefit \cite{6587554}. Also, without subcarrier selection, the reliability of the systems will be much lower than that with our proposed adaptive scheme.

\subsubsection{FPSK}
As a well known modulation scheme \cite{7763523}, we also take FPSK as a comparison benchmark in this paper. In FPSK, only one subcarrier will be activated and the index of the active subcarrier is used for conveying information. The transmission rate is given by
\begin{equation}\small
B_{FPSK}=\log_2(M)+\lfloor\log_2(N_T)\rfloor~~~~~~~~[\mathrm{bpcu}].
\end{equation}

\begin{figure*}[!t]
\centering
\includegraphics[width=5.5in]{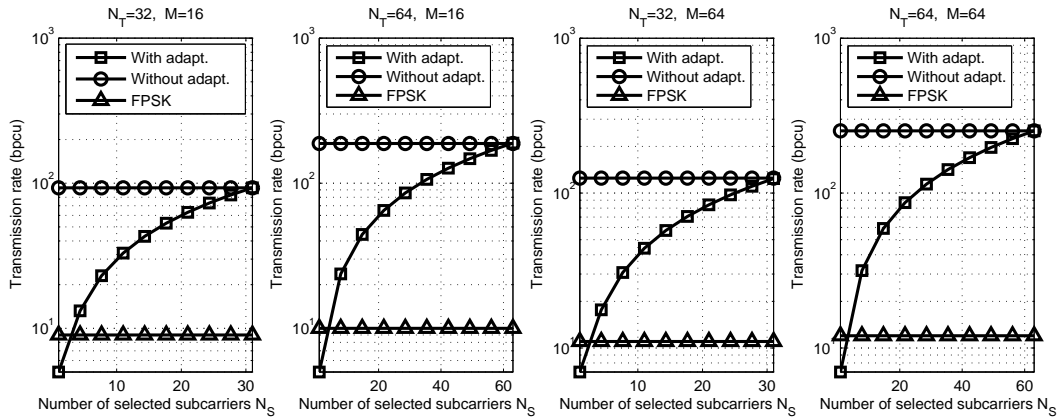}
\caption{Transmission rates of different schemes vs. $N_S$.}
\label{fig_transmission_rate}
\end{figure*}

The FPSK scheme would lead to a low-complexity system, since the detection process is much simpler than our proposed adaptive OFDM IM scheme and the zero-active subcarrier dilemma does not exist either. On the other hand, the transmission rate of the systems using FPSK might be significantly lower than that of systems using our proposed schemes, because the spectral efficiency of FPSK in frequency dimension is too low by activating only a single subcarrier in each instant. Detailed numerical comparisons in terms of outage performance, network capacity and error performance for the adaptive OFDM IM scheme and these two benchmarks will be given in Section \ref{nr}. The transmission rates of these three schemes are plotted in Fig. \ref{fig_transmission_rate} for illustration purposes.

\section{Outage Performance Analysis}\label{opa}
\subsection{Decentralized mapping scheme selection}
\subsubsection{Complementary transmission mode}
When $\mathbf{b}_S(1)=[0,0,\dots,0]$ is required to be transmitted, the complementary transmission mode will be activated and the standby subcarrier $\tilde{n}_i$ will be employed to convey the APM symbol $\chi_{m_{\tilde{n}}}$. By (\ref{xuanzejizhibuchong}), the standby subcarrier $\tilde{n}_i$ is the $(N_T-N_S)$th worst among all $N_T$ subcarriers and the only one active subcarrier in the complementary transmission mode. Therefore, the outage probability regarding the subcarrier $\tilde{n}_i$ in the $i$th hop is given by
\begin{equation}\small
P_{o-i}(s|k=1)=F_{i\left(N_T-N_S\right)}\left(\frac{sN_0}{P_t}\right),
\end{equation}
where
\begin{equation}\label{dsa411232dsworstous}\small
F_{i\left(\xi\right)}\left(s\right)=\sum_{n=\xi}^{N_T}\binom{N_T}{n}\left(F_{i}\left(s\right)\right)^n\left(1-F_{i}\left(s\right)\right)^{N_T-n}
\end{equation}
denotes the outage probability of the $\xi$th order statistic of the channel gain among $N_T$ subcarriers in the $i$th hop \cite{6678968}. Hence, the outage probability of the complementary transmission mode is
\begin{equation}\small
\begin{split}
P_{o}(s|k=1)=&P_{o-1}(s|k=1)+P_{o-2}(s|k=1)-P_{o-1}(s|k=1)P_{o-2}(s|k=1).
\end{split}
\end{equation}

\subsubsection{Analysis of average outage probability}
In order to analyze the outage performance when $k\neq 1$, we propose and prove a lemma as follows.
\begin{lemma}\label{shjdakh2xmznkjca1leemas1}
For $k\neq 1$, the generic conditional outage probability for the $i$th hop can be determined by
\begin{equation}\label{averagewetouts}\small
P_{o-i}(s|k)=\sum_{\xi=N_T-N_S+1}^{N_T-N_A(k)+1}\Upsilon(k,\xi)F_{i\left(\xi\right)}\left(\frac{sN_0N_A(k)}{P_t}\right),
\end{equation}
where $\Upsilon(k,\xi)=\left.\binom{N_T-\xi}{N_A(k)-1} \middle / \binom{N_S}{N_A(k)} \right.$.
\end{lemma}
\begin{IEEEproof}
See Appendix \ref{proofconditionalproba}
\end{IEEEproof}
Due to the DF forwarding protocol, the conditional outage probability considering two hops is derived by
\begin{equation}\small
P_{o}(s|k)=P_{o-1}(s|k)+P_{o-2}(s|k)-P_{o-1}(s|k)P_{o-2}(s|k).
\end{equation}
Because $\mathbf{b}_S(k)$ is equiprobable, the average outage probability is given by
\begin{equation}\label{exactoutagepoesdsaj2tt1}\small
\bar P_{o}(s)=\frac{1}{2^{N_S}}\sum_{k\in\mathcal{K}}P_{o}(s|k).
\end{equation}
By employing a power series expansion at $P_t/N_0\rightarrow \infty$, we can obtain the asymptotic expression of $\bar P_{o}(s)$ as
\begin{equation}\label{asymptoticexpressout1}\small
\begin{split}
\bar P_{o}(s)&\sim \frac{\binom{N_T}{N_T-N_S}}{2^{N_S}}\left[\left(\frac{1}{\mu_1}\right)^{N_T-N_S}+\left(\frac{1}{\mu_2}\right)^{N_T-N_S}\right]\left(\frac{sN_0}{P_t}\right)^{N_T-N_S},
\end{split}
\end{equation}
by which we can find the diversity order to be $N_T-N_S$.

\subsection{Centralized mapping scheme selection}\label{dsadkasji2secsouts2}
\subsubsection{Complementary transmission mode}
Because the mapping schemes used at the source and relay are the same by the centralized mapping scheme selection, we have to integrate the channels in two hops into a whole, which is termed a \textit{link}. When DF forwarding protocol is adopted, due to the bottleneck effect, the PDF and CDF of the gain regarding a certain unsorted link can be determined from (\ref{channelpdfcdf}) to be
\begin{equation}\label{kjdlk1dsad}\small
\phi(s)=\mathrm{exp}\left(-s/\mu_{\Sigma}\right)/\mu_{\Sigma}~~\Leftrightarrow~~\Phi(s)=1-\mathrm{exp}\left(-s/\mu_{\Sigma}\right),
\end{equation}
where $\mu_{\Sigma}={\mu_1\mu_2}/{(\mu_1+\mu_2)}$.

In a similar manner to the decentralized case, we can have the outage probability of the complementary transmission mode as
\begin{equation}\small
P_o(s|k=1)=\Phi_{\left(N_T-N_S\right)}\left(\frac{sN_0}{P_t}\right),
\end{equation}
where
\begin{equation}\small
\Phi_{\left(\xi\right)}\left(s\right)=\sum_{n=\xi}^{N_T}\binom{N_T}{n}\left(\Phi\left(s\right)\right)^n\left(1-\Phi\left(s\right)\right)^{N_T-n}
\end{equation}
denotes the outage probability of the $\xi$th order statistic of the link gain among $N_T$ links.

\subsubsection{Analysis of average outage probability}
Now, let us generalize the analysis to an arbitrary case when $k\neq 1$, and similarly obtain the generic conditional average outage probability by
\begin{equation}\small
P_o(s|k)=\sum_{\xi=N_T-N_S+1}^{N_T-N_A(k)+1}\Upsilon(k,\xi)\Phi_{\left(\xi\right)}\left(\frac{sN_0N_A(k)}{P_t}\right).
\end{equation}

Then, by (\ref{2d1sa23142}) and the assumption of equiprobable $\mathbf{b}_S(k)$, the average outage probability can be calculated by (\ref{exactoutagepoesdsaj2tt1}) as well. By employing a power series expansion at $P_t/N_0\rightarrow \infty$, we can obtain the asymptotic expression of $\bar P_{o}(s)$ as
\begin{equation}\label{asymptoticexpressout2}\small
\bar P_{o}(s)\sim \frac{1}{2^{N_S}}\binom{N_T}{N_T-N_S}\left(\frac{sN_0}{\mu_{\Sigma}P_t}\right)^{N_T-N_S},
\end{equation}
by which we can find the diversity order to be $N_T-N_S$.

\section{Network Capacity Analysis}\label{anca}
\subsection{Decentralized mapping scheme selection}
\subsubsection{Complementary transmission mode}
By (\ref{dsadoutsctutasn}), we can reduce the expression of the network capacity of the complementary transmission mode to
\begin{equation}\label{dskj2dsx23cdacaps2s62}\small
C(1)=\frac{1}{2}\log_2\left(1+\min\left\lbrace\gamma_1(1,\tilde{n}_1),\gamma_2(1,\tilde{n}_2)\right\rbrace\right),
\end{equation}
in which the PDF of $\min\left\lbrace\gamma_1(1,\tilde{n}_1),\gamma_2(1,\tilde{n}_2)\right\rbrace$ after performing mapping scheme selections can be expressed as $\psi_{\left(N_T-N_S,N_T-N_S\right)}\left(\frac{sN_0}{P_t}\right)$ and 
\begin{equation}\label{dsadas4552dscouspds}\small
\begin{split}
\psi_{\left(\xi,\eta\right)}\left(xs\right)&=xf_{1\left(\xi\right)}\left(xs\right)\left(1-F_{2\left(\eta\right)}\left(xs\right)\right)+xf_{2\left(\eta\right)}\left(xs\right)\left(1-F_{1\left(\xi\right)}\left(xs\right)\right),
\end{split}
\end{equation}
where
\begin{equation}\small
f_{i(\xi)}(s)=\frac{\mathrm{d}F_{i(\xi)}(s)}{\mathrm{d}s}=\frac{N_T!\left(F_{i}(s)\right)^{\xi-1}\left(1-F_{i}(s)\right)^{N_T-\xi}f_{i}(s)}{(\xi-1)!(N_T-\xi)!}.
\end{equation}

Therefore, the average capacity over $h_1(\tilde{n}_1)$ and $h_2(\tilde{n}_2)$, for the complementary transmission mode can be obtained as
\begin{equation}\small
\begin{split}
\bar{C}(1)&=\underset{h_1(\tilde{n}_1),h_2(\tilde{n}_2)}{\mathbb{E}}\left\lbrace C(1)\right\rbrace=\int_{0}^{\infty}\frac{1}{2}\log_2\left(1+s\right)\psi_{\left(N_T-N_S,N_T-N_S\right)}\left(\frac{sN_0}{P_t}\right)\mathrm{d}s=\Lambda_G\left(N_T-N_S,N_T-N_S,\frac{N_0}{P_t}\right).
\end{split}
\end{equation}
Here, $\Lambda_G(\xi,\eta,x)$ is a defined special function, which can be explicitly written in (\ref{ejsckaL99mbscs235}). The expressions $\lambda_i(\xi,x)$ and $\nu_{i,j}(\xi,\eta,x)$ can be derived in closed form in (\ref{lambdaexpression}) and (\ref{nuexpression}), where $(i,j)\in\{(1,2),(2,1)\}$; $\mathrm{Ei}(x)=-\int_{-x}^{\infty}\frac{e^{-u}}{u}\mathrm{d}u$ is the exponential integral function and $\kappa_{i,j}(\xi,n,d,b)=\frac{N_T-\xi+1+b}{\mu_i}+\frac{N_T+d-n}{\mu_j}$ represents an intermediate coefficient.

\begin{figure*}[t!]
\begin{equation}\label{ejsckaL99mbscs235}\small
\begin{split}
\Lambda_G(\xi,\eta,x)&=\underbrace{\int_{0}^{\infty}\frac{x}{2}\log_2(1+s)f_{1(\xi)}(xs)\mathrm{d}s}_{\lambda_1(\xi,x)}-\underbrace{\int_{0}^{\infty}\frac{x}{2}\log_2(1+s)f_{1(\xi)}(xs)F_{2(\eta)}(xs)\mathrm{d}s}_{\nu_{1,2}(\xi,\eta,x)}\\
&+\underbrace{\int_{0}^{\infty}\frac{x}{2}\log_2(1+s)f_{2(\eta)}(xs)\mathrm{d}s}_{\lambda_2(\eta,x)}-\underbrace{\int_{0}^{\infty}\frac{x}{2}\log_2(1+s)f_{2(\eta)}(xs)F_{1(\xi)}(xs)\mathrm{d}s}_{\nu_{2,1}(\eta,\xi,x)}
\end{split}
\end{equation}
\hrule
\end{figure*}

\begin{figure*}[t!]
\begin{equation}\label{lambdaexpression}\small
\lambda_i(\xi,x)=\frac{N_T!}{(2\ln 2)(\xi-1)!(N_T-\xi)!}\sum_{b=0}^{\xi-1}\left[\binom{\xi-1}{b}(-1)^{b+1}\mathrm{exp}\left(\frac{x}{\mu_i}\left(N_T-\xi+1+b\right)\right)\mathrm{Ei}\left(-(N_T-\xi+1+b)\right)\right]
\end{equation}
\hrule
\end{figure*}

\begin{figure*}[t!]
\begin{equation}\label{nuexpression}\small
\begin{split}
\nu_{i,j}(\xi,\eta,x)&=\frac{N_T!}{\mu_i(2\ln 2) (\xi-1)!(N_T-\xi)!}\sum_{n=\eta}^{N_T}\sum_{d=0}^n\sum_{b=0}^{\xi-1}\binom{N_T}{n}\binom{n}{d}\binom{\xi-1}{b}\\
&~~~~~~~~~\times(-1)^{b+d+1}\frac{\mathrm{exp}\left(x\kappa_{i,j}(\xi,n,d,b)\right)\mathrm{Ei}\left(-x\kappa_{i,j}(\xi,n,d,b)\right)}{\kappa_{i,j}(\xi,n,d,b)}
\end{split}
\end{equation}
\hrule
\end{figure*}

\subsubsection{Analysis of average network capacity}
For $k\neq 1$, there will be $N_A(k)$ active subcarriers at each hop, whose orders are denoted by $\mathbf{\Xi}_1(k)=\{\xi_1,\xi_2,\dots,\xi_{N_A(k)}\}$ and $\mathbf{\Xi}_2(k)=\{\eta_1,\eta_2,\dots,\eta_{N_A(k)}\}$, where $\xi_i\neq\xi_j$ and $\eta_i\neq\eta_j$ for $i\neq j$ and $N_T-N_S+1\leq\xi_i,\eta_i\leq N_T$, $\forall~1 \leq i\leq N_A(k)$. The set of permutations of $\mathbf{\Xi}_1(k)$ and $\mathbf{\Xi}_2(k)$ is denoted as $\mathcal{D}(k)$ and $\mathrm{Card}(\mathcal{D}(k))={N_S!}/{(N_S-N_A(k))!}$. Therefore, in order to determine the conditional average capacity over $\mathbf{H}_1(\hat{c}_1)$ and $\mathbf{H}_2(\hat{c}_2)$, we propose and prove a lemma as follows.
\begin{lemma}\label{liemadier2}
For $k\neq 1$, the conditional average capacity over $\mathbf{H}_1(\hat{c}_1)$ and $\mathbf{H}_2(\hat{c}_2)$ is given by
\begin{equation}\label{capacituygdsak2s}\small
\begin{split}
\bar{C}(k)&=\underset{\mathbf{H}_1(\hat{c}_1),\mathbf{H}_2(\hat{c}_2)}{\mathbb{E}}\left\lbrace C(k)\right\rbrace=\sum_{\mathbf{\Xi}_1(k)\in\mathcal{D}(k)}\sum_{\mathbf{\Xi}_2(k)\in\mathcal{D}(k)}\left(\frac{(N_S-N_A(k))!}{N_S!}\right)^2\underset{n\in\mathcal{N}_A(k)}{\sum}\Lambda_G\left(\xi_n,\eta_n,\frac{N_0N_A(k)}{P_t}\right).
\end{split}
\end{equation}
\end{lemma}
\begin{IEEEproof}
See Appendix \ref{proofsaikdjakcapa}.
\end{IEEEproof}

As a result of the equiprobable assumption of $\mathbf{b}_S(k)$, the average network capacity can be obtained by
\begin{equation}\label{averagenetowrkscasss1}\small
\begin{split}
\bar{C}&=\underset{k\in\mathcal{K}}{\mathbb{E}}\left\lbrace \bar{C}(k)\right\rbrace=\frac{1}{2^{N_S}}\sum_{k\in\mathcal{K}}\bar{C}(k).
\end{split}
\end{equation}

\subsection{Centralized mapping scheme selection}
\subsubsection{Complementary transmission mode}
We can express the network capacity of the complementary mode by modifying (\ref{dskj2dsx23cdacaps2s62}) and determine the PDF of $\min\left\lbrace\gamma_1(1,\tilde{n}),\gamma_2(1,\tilde{n})\right\rbrace$ by $\phi_{\left(N_T-N_S\right)}\left(\frac{sN_0}{P_t}\right)$, where
\begin{equation}\small
\begin{split}
\phi_{\left(\xi\right)}\left(xs\right)&=\frac{\mathrm{d}\Phi_{\left(\xi\right)}\left(xs\right)}{\mathrm{d}s}=\frac{x(N_T)!(\Phi(xs))^{\xi-1}(1-\Phi(xs))^{N_T-\xi}\phi(xs)}{(\xi-1)!(N_T-\xi)!}.
\end{split}
\end{equation}

Subsequently, the average capacity of the complementary mode can be determined by
\begin{equation}\small
\begin{split}
&\bar{C}(1)=\underset{h_1(\tilde{n}),h_2(\tilde{n})}{\mathbb{E}}\left\lbrace C(1)\right\rbrace=\int_{0}^{\infty}\frac{1}{2}\log_2\left(1+s\right)\phi_{\left(N_T-N_S\right)}\left(\frac{sN_0}{P_t}\right)\mathrm{d}s=\Lambda_L\left(N_T-N_S,\frac{N_0}{P_t}\right),
\end{split}
\end{equation}
where $\Lambda_L(\xi,x)$ can be similarly defined as (\ref{lambdaexpression}) by replacing $\mu_i$ with $\mu_\Sigma$.

\subsubsection{Analysis of average network capacity}
For $k\neq 1$, we can derive the conditional average capacity similarly as for the decentralized case by
\begin{equation}\label{adoi298djkmn2}\small
\begin{split}
\bar{C}(k)&=\underset{\mathbf{H}_1(\hat{c}),\mathbf{H}_2(\hat{c})}{\mathbb{E}}\left\lbrace C(k)\right\rbrace=\sum_{\mathbf{\Xi}(k)\in\mathcal{D}(k)}\left(\frac{(N_S-N_A(k))!}{N_S!}\right)\underset{\begin{subarray}~n\in\mathcal{N}_A(k)\end{subarray}}{\sum}\Lambda_L\left(\xi_n,\frac{N_0N_A(k)}{P_t}\right).
\end{split}
\end{equation}
Then, by (\ref{averagenetowrkscasss1}), the average network capacity for the centralized case can be derived.

\section{Error Performance Analysis}\label{epa}
\subsection{Decentralized mapping scheme selection}
According to \cite{789668}, we can determine the conditional SER in the $i$th hop as
\begin{equation}\label{dsa524sajkjkpeis}\small
\begin{split}
&P_{e-i}(\dot{\mathbf{X}}(\dot{k})\rightarrow \hat{\mathbf{X}}(\hat{k})\vert \mathbf{H}_i(\hat{c}_i),h_i(\tilde{n}_i))=Q\left(\sqrt{\frac{P_t}{N_0}\begin{Vmatrix}\langle\mathbf{H}_i(\hat{c}_i),h_i(\tilde{n}_i)\rangle\left(\frac{\dot{\mathbf{X}}(\dot{k})}{\sqrt{N_A(\dot{k})}}-\frac{\hat{\mathbf{X}}(\hat{k})}{\sqrt{N_A(\hat{k})}}\right)\end{Vmatrix}_F^2}\right),
\end{split}
\end{equation}
where $Q(x)=\frac{1}{\sqrt{2\pi}}\int_{x}^{\infty}\mathrm{exp}\left(-u^2/2\right)\mathrm{d}u$ is the Q-function. Subsequently, in order to remove the conditions on $\mathbf{H}_i(\hat{c}_i)$, $h_{i}(\tilde{n}_i)$ and obtain an accurate closed-form approximation, we first need to approximate the Q-function by \cite{1188428}

\begin{equation}\label{dsakdjsakjh2cxqhanshu}\small
Q(x)\approx \frac{1}{12}\mathrm{exp}\left(-\frac{x^2}{2}\right)+\frac{1}{4}\mathrm{exp}\left(-\frac{2x^2}{3}\right),
\end{equation}
which becomes accurate when $x$ is large, and has been proved to be effective in obtaining the approximate SER for traditional OFDM IM as shown in \cite{6587554}. Also, we define a function $\Theta_{G_i}\left(\dot{\mathbf{X}}(\dot{k}),\hat{\mathbf{X}}(\hat{k}),\mathbf{\Xi}_i\right)$ and simplify it in (\ref{defineyigexinhanshu}), where $\mathbf{\Xi}_i\in\mathcal{D}$ is the vector of ordered indices of the channel gains regarding $N_S$ selected subcarriers in the $i$th hop and $\mathcal{D}$ is the permutation set; $\Gamma(x)=\int_{0}^{\infty}u^{x-1}\mathrm{exp}(-u)\mathrm{d}u$ is the gamma function.

\begin{figure*}[t!]
\begin{equation}\label{defineyigexinhanshu}\small
\begin{split}
&\Theta_{G_i}\left(\dot{\mathbf{X}}(\dot{k}),\hat{\mathbf{X}}(\hat{k}),\mathbf{\Xi}_i\right)\\
&=\underbrace{\int_{0}^{\infty}\int_{0}^{\infty}\dots\int_{0}^{\infty}}_{(N_S+1)-\mathrm{fold}}Q\left(\sqrt{\frac{P_t}{N_0}\begin{Vmatrix}
\mathrm{diag}\{u_1,u_2,\dots,u_{N_S+1}\}\left(\frac{\dot{\mathbf{X}}(\dot{k})}{\sqrt{N_A(\dot{k})}}-\frac{\hat{\mathbf{X}}(\hat{k})}{\sqrt{N_A(\hat{k})}}\right)
\end{Vmatrix}_F^2}\right)\\
&\times\left(f_{i(N_T-N_S)}(u_{N_S+1})\prod_{n=1}^{N_S}f_{i(\xi_n)}(u_n)\right)\mathrm{d}u_1\mathrm{d}u_2\dots\mathrm{d}u_{N_S+1}\\
&\approx\frac{1}{12}\left(\frac{N_T!\Gamma\left(N_S+1+\frac{\mu_iP_t}{2N_0}\left\lvert\dot{\chi}_{\dot{m}_{\tilde{n}}}-\hat{\chi}_{\hat{m}_{\tilde{n}}}\right\rvert^2\right)}{N_S!\Gamma\left(N_T+1+\frac{\mu_iP_t}{2N_0}\left\lvert\dot{\chi}_{\dot{m}_{\tilde{n}}}-\hat{\chi}_{\hat{m}_{\tilde{n}}}\right\rvert^2\right)}\right)\prod_{n=1}^{N_S}\left(\frac{N_T!\Gamma\left(N_T-\xi_n+1+\frac{\mu_iP_t}{2N_0}\left\lvert\frac{\dot{x}(\dot{m}_n,n)}{\sqrt{N_A(\dot{k})}}-\frac{\hat{x}(\hat{m}_n,n)}{\sqrt{N_A(\hat{k})}}\right\rvert^2\right)}{(N_T-\xi_n)!\Gamma\left(N_T+1+\frac{\mu_iP_t}{2N_0}\left\lvert\frac{\dot{x}(\dot{m}_n,n)}{\sqrt{N_A(\dot{k})}}-\frac{\hat{x}(\hat{m}_n,n)}{\sqrt{N_A(\hat{k})}}\right\rvert^2\right)}\right)\\
&~~~~+\frac{1}{4}\left(\frac{N_T!\Gamma\left(N_S+1+\frac{2\mu_iP_t}{3N_0}\left\lvert\dot{\chi}_{\dot{m}_{\tilde{n}}}-\hat{\chi}_{\hat{m}_{\tilde{n}}}\right\rvert^2\right)}{N_S!\Gamma\left(N_T+1+\frac{2\mu_iP_t}{3N_0}\left\lvert\dot{\chi}_{\dot{m}_{\tilde{n}}}-\hat{\chi}_{\hat{m}_{\tilde{n}}}\right\rvert^2\right)}\right)\prod_{n=1}^{N_S}\left(\frac{N_T!\Gamma\left(N_T-\xi_n+1+\frac{2\mu_iP_t}{3N_0}\left\lvert\frac{\dot{x}(\dot{m}_n,n)}{\sqrt{N_A(\dot{k})}}-\frac{\hat{x}(\hat{m}_n,n)}{\sqrt{N_A(\hat{k})}}\right\rvert^2\right)}{(N_T-\xi_n)!\Gamma\left(N_T+1+\frac{2\mu_iP_t}{3N_0}\left\lvert\frac{\dot{x}(\dot{m}_n,n)}{\sqrt{N_A(\dot{k})}}-\frac{\hat{x}(\hat{m}_n,n)}{\sqrt{N_A(\hat{k})}}\right\rvert^2\right)}\right)
\end{split}
\end{equation}
\hrule
\end{figure*}

By (\ref{shijiugongset1}), (\ref{shijiugongset2}) and (\ref{shijiugongset3}), we can obtain the unconditional end-to-end SER by
\begin{equation}\small
\begin{split}
&P_{e}(\dot{\mathbf{X}}(\dot{k}))=\Omega_{G_1}(\dot{\mathbf{X}}(\dot{k}))+\Omega_{G_2}(\dot{\mathbf{X}}(\dot{k}))-\Omega_{G_1}(\dot{\mathbf{X}}(\dot{k}))\Omega_{G_2}(\dot{\mathbf{X}}(\dot{k})),
\end{split}
\end{equation}
where $\Omega_{G_i}(\dot{\mathbf{X}}(\dot{k}))$ is defined as follows:
\begin{equation}\label{omegahanshu}\small
\begin{split}
\Omega_{G_i}(\dot{\mathbf{X}}(\dot{k}))=\frac{1}{N_S!}\underset{\hat{\mathbf{X}}(\hat{k})\neq\dot{\mathbf{X}}(\dot{k})}{\sum}\underset{\mathbf{\Xi}_i\in\mathcal{D}}{\sum}\Theta_{G_i}\left(\dot{\mathbf{X}}(\dot{k}),\hat{\mathbf{X}}(\hat{k}),\mathbf{\Xi}_i\right).
\end{split}
\end{equation}

Therefore, by (\ref{SERzuizhongban}), the average SER can be determined by
\begin{equation}\label{dsakdsajkdhk1dsaverageser}\small
\bar P_{e}=\underset{\dot{\mathbf{X}}(\dot{k})\in\mathcal{X}}{\sum} \varphi(\dot{\mathbf{X}}(\dot{k})) P_{e}(\dot{\mathbf{X}}(\dot{k})),
\end{equation}
where $\varphi(\dot{\mathbf{X}}(\dot{k}))$ represents the probability that $\dot{\mathbf{X}}(\dot{k})$ is transmitted, and can be written as
\begin{equation}\small
\varphi(\dot{\mathbf{X}}(\dot{k}))=\left. \binom{N_S}{N_A(k)} \middle / \left(2^{N_S} M^{\max\{1,N_A(k)\}}\right) \right..
\end{equation}

\subsection{Centralized mapping scheme selection}
Following the concept of a link as explained in Section \ref{dsadkasji2secsouts2},  we can let
\begin{equation}\small
\begin{split}
&\mathbf{L}(\hat{c})=\mathrm{diag}\{l(\hat{c},1),l(\hat{c},2),\dots,l(\hat{c},N_S)\}\\
&=\mathrm{diag}\{\min\{h_1(\hat{c},1),h_2(\hat{c},1)\},\min\{h_1(\hat{c},2),h_2(\hat{c},2)\},\dots,\min\{h_1(\hat{c},N_S),h_2(\hat{c},N_S)\}\},
\end{split}
\end{equation}
and
\begin{equation}\small
l(\tilde{n})=\min\{h_1(\tilde{n}),h_2(\tilde{n})\}.
\end{equation}
Then, when applying the centralized mapping scheme selection, we can similarly obtain the unconditional end-to-end SER by
\begin{equation}\label{465d4saeidererrser}\small
\begin{split}
&P_{e}(\dot{\mathbf{X}}(\dot{k}))=\Omega_L(\dot{\mathbf{X}}(\dot{k})),
\end{split}
\end{equation}
where
\begin{equation}\label{tangjisndsajkh2ssousaoes}\small
\begin{split}
\Omega_L(\dot{\mathbf{X}}(\dot{k}))=\frac{1}{N_S!}{\underset{\hat{\mathbf{X}}(\hat{k})\neq\dot{\mathbf{X}}(\dot{k})}{\sum}\underset{\mathbf{\Xi}\in\mathcal{D}}{\sum}\Theta_L\left(\dot{\mathbf{X}}(\dot{k}),\hat{\mathbf{X}}(\hat{k}),\mathbf{\Xi}\right)}
\end{split}
\end{equation}
and $\Theta_L\left(\dot{\mathbf{X}}(\dot{k}),\hat{\mathbf{X}}(\hat{k}),\mathbf{\Xi}\right)$ can be similarly defined and approximated as (\ref{defineyigexinhanshu}) by replacing $\mu_i$ with $\mu_{\Sigma}$. As a result, substituting (\ref{465d4saeidererrser}) into (\ref{dsakdsajkdhk1dsaverageser}) yields the average SER for the centralized mapping scheme selection case.

\begin{figure*}[t!]
    \centering
    \begin{subfigure}[t]{0.5\textwidth}
        \centering
        \includegraphics[width=3.5in]{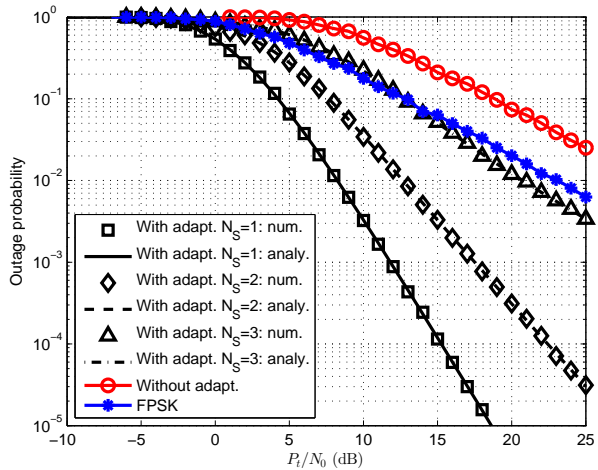}
        \caption{Decentralized mapping scheme selection: $N_T=4$.}
    \end{subfigure}%
~
    \begin{subfigure}[t]{0.5\textwidth}
        \centering
        \includegraphics[width=3.5in]{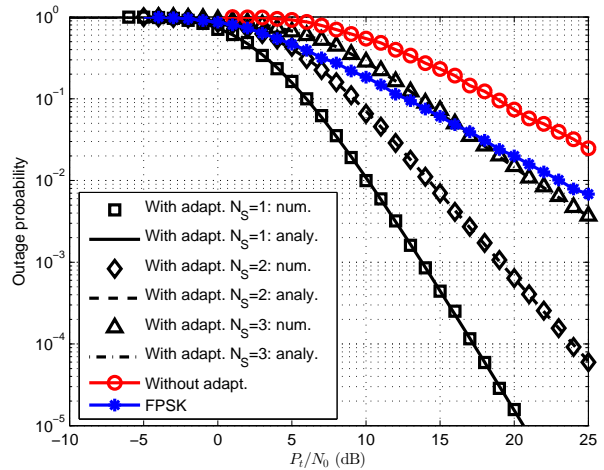}
        \caption{Centralized mapping scheme selection: $N_T=4$.}
    \end{subfigure}
~
    \centering
    \begin{subfigure}[t]{0.5\textwidth}
        \centering
        \includegraphics[width=3.5in]{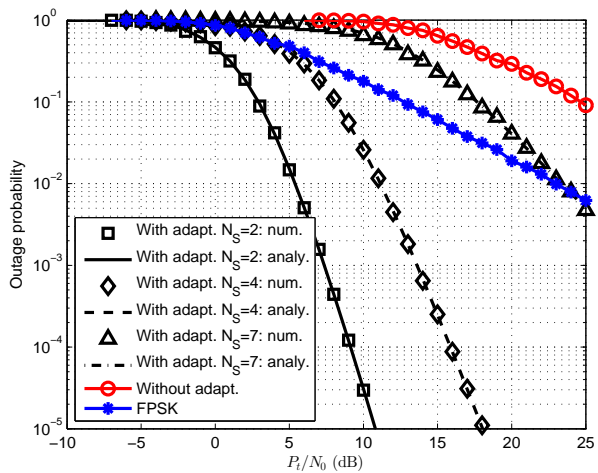}
        \caption{Decentralized mapping scheme selection: $N_T=8$.}
    \end{subfigure}%
~
    \begin{subfigure}[t]{0.5\textwidth}
        \centering
        \includegraphics[width=3.5in]{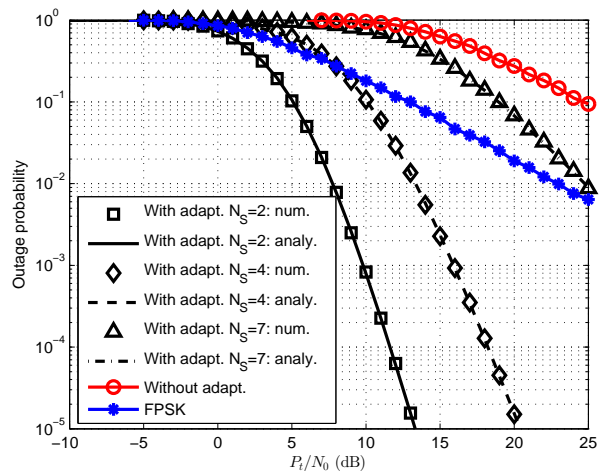}
        \caption{Centralized mapping scheme selection: $N_T=8$.}
    \end{subfigure}
    \caption{Average outage probability vs. ratio of transmit power to noise power $P_t/N_0$ for different schemes and $N_T$.}
    \label{fig_outage_probability_vs_pt}
\end{figure*}

\section{Numerical Results}\label{nr}
To verify the analysis given in Section \ref{opa}, Section \ref{anca} and Section \ref{epa}, we normalize $\mu_1=\mu_2=1$, $s=1$ as well as $N_0=1$ and then carry out the simulations with different $N_T$ and $N_S$ as well as $M$ regarding average outage probability, network capacity and SER, respectively. The simulation results are shown and discussed as follows.

\subsection{Simulations of outage performance}
The relations between average outage probability and the ratio of transmit power to noise power $P_t/N_0$ for different transmission schemes with different $N_T$ were numerically simulated by Monte Carlo methods and are shown in Fig. \ref{fig_outage_probability_vs_pt}. From this figure, the numerical results match the analytical results, which verifies our analysis in Section \ref{opa}. Meanwhile, from these two figures, some important features of adaptive OFDM IM are revealed. First, the systems with decentralized and centralized mapping scheme selections share the same diversity order, but have different coding gains. This aligns with our analytical results given in (\ref{asymptoticexpressout1}) and (\ref{asymptoticexpressout2}).

Second, our proposed schemes always have a better outage performance compared to the scheme without adaptation, and this performance advantage is brought by reducing the number of subcarriers used for OFDM IM. Besides, our proposed schemes with any $N_S$ will enjoy a better outage performance than that of FPSK, as long as $P_t/N_0$ is relatively large. Also, a large $N_T$ will lead to a better outage performance, since the adaptive OFDM IM will have more choices when selecting the mapping scheme. On the other hand, a larger $N_S$ will result in a larger average number of active subcarriers, which will degrade the outage performance, as all active subcarriers need to be ensured not in outage as specified in \textit{Definition \ref{defoutages323x}}. Therefore, when $N_S$ approaches $N_T$, the outage performance of the adaptive schemes will be close to that of the scheme without adaptation. This reflects a diversity-multiplexing trade-off in terms of $N_S$ in adaptive OFDM IM, which is worth further investigating as a future work\footnote{A similar trade-off for MIMO systems has been found and analyzed in \cite{1197843}.}.

\begin{figure}[!t]
\centering
\includegraphics[width=5.5in]{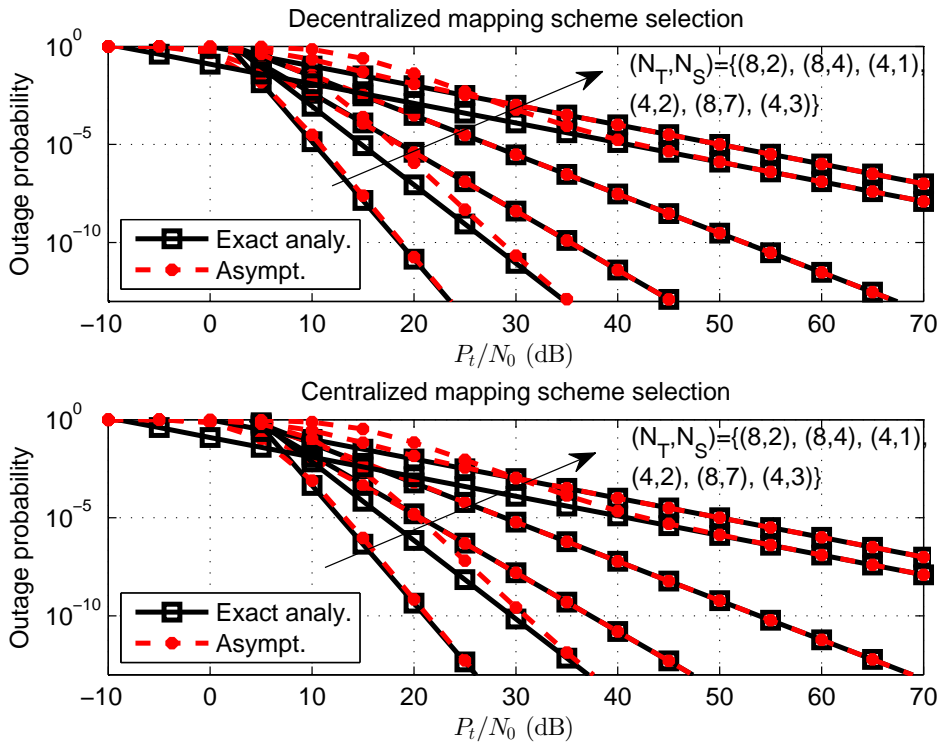}
\caption{Exact expressions and asymptotic expressions of outage probability for decentralized and centralized mapping scheme selections.}
\label{fig_outage_probability_asympt}
\end{figure}

Furthermore, we have plotted the average outage probabilities and the asymptotic expressions at high SNR for both decentralized and centralized cases in Fig. \ref{fig_outage_probability_asympt}, which further verifies the derived asymptotic relations in Section \ref{opa} and the diversity discussion above.

\begin{figure*}[t!]
    \centering
    \begin{subfigure}[t]{0.5\textwidth}
        \centering
        \includegraphics[width=3.5in]{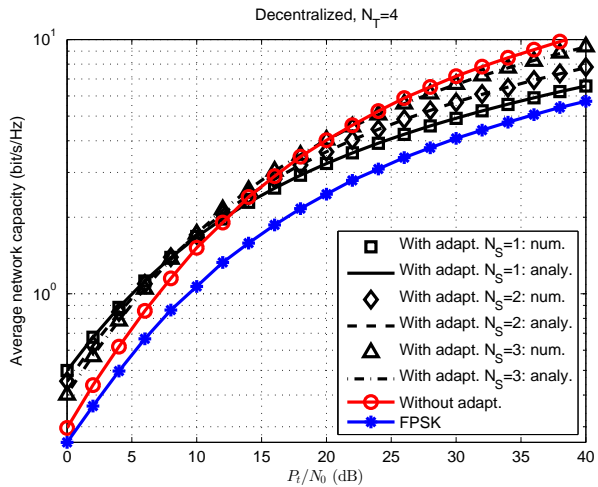}
        \caption{Decentralized mapping scheme selection: $N_T=4$.}
    \end{subfigure}%
~
    \begin{subfigure}[t]{0.5\textwidth}
        \centering
        \includegraphics[width=3.5in]{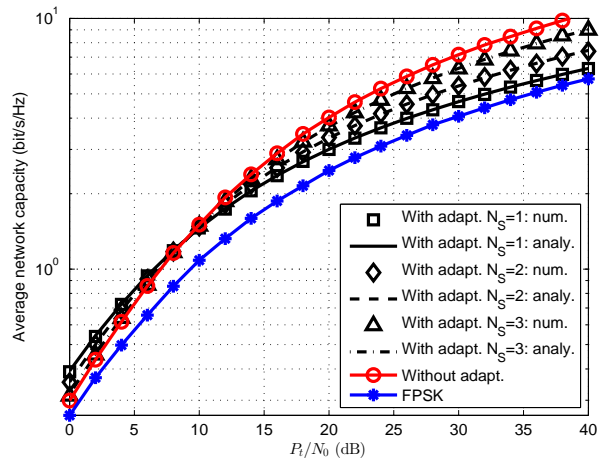}
        \caption{Centralized mapping scheme selection: $N_T=4$.}
    \end{subfigure}
~
    \centering
    \begin{subfigure}[t]{0.5\textwidth}
        \centering
        \includegraphics[width=3.5in]{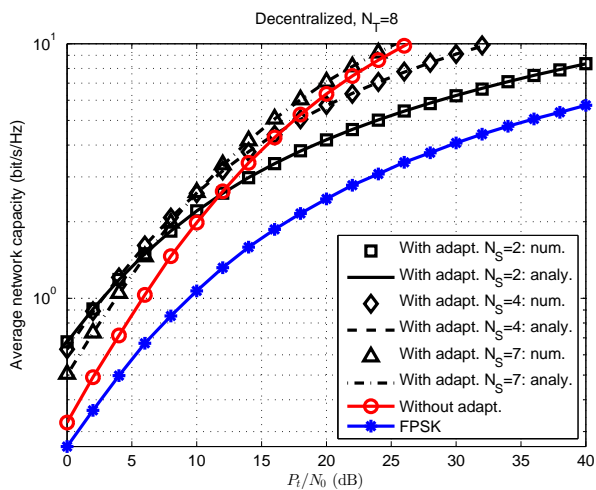}
        \caption{Decentralized mapping scheme selection: $N_T=8$.}
    \end{subfigure}%
~
    \begin{subfigure}[t]{0.5\textwidth}
        \centering
        \includegraphics[width=3.5in]{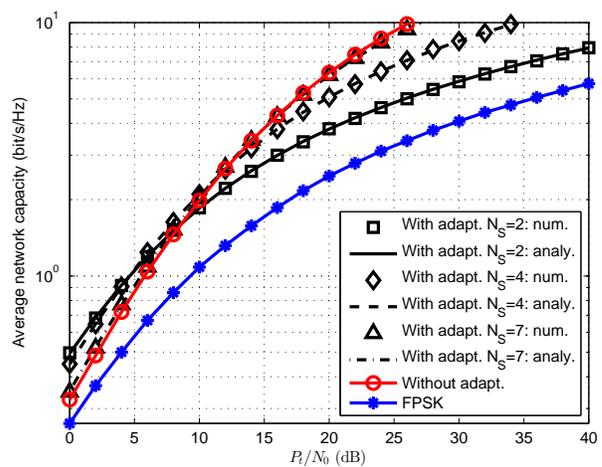}
        \caption{Centralized mapping scheme selection: $N_T=8$.}
    \end{subfigure}
    \caption{Average network capacity vs. ratio of transmit power to noise power $P_t/N_0$ for different schemes and $N_T$.}
    \label{fig_capacity_vs_pt}
\end{figure*}

\subsection{Simulations of network capacity}
By taking the same settings for the simulations of outage probability, we carried out Monte Carlo simulations to study the average network capacity for different transmission schemes and present the results in Fig. \ref{fig_capacity_vs_pt}. The numerical results verify our analysis in Section \ref{anca}. Therefore, they can be used as effective tools to estimate the average network capacity of systems applying adaptive OFDM IM in two-hop networks. Meanwhile, it is expected that the system using the decentralized mapping scheme selection always outperforms the system using the centralized mapping scheme selection in terms of average network capacity. Also, a higher number of subcarriers $N_T$ will lead to a higher average network capacity, since the average number of active subcarriers becomes higher and there are more options for mapping schemes selections. However, when it comes to the effects of $N_S$, the situation becomes complicated and interesting. On the one hand, with a small $P_t/N_0$, a smaller $N_S$ will lead to a higher average network capacity, since the power allocated in each subcarrier matters and has a significant impact on the capacity. Therefore, when $N_S$ is small, the average transmit power allocated to each active subcarrier will be large, which yields a positive impact on the network capacity. On the other hand, when $P_t/N_0$ is large, because of the logarithm function (c.f. (\ref{dsadoutsctutasn})), the power allocated in each subcarrier only has a small impact on the capacity, while the number of terms taken in the summation operation matters. Therefore, as long as $P_t/N_0$ is sufficiently large, a larger $N_S$ will lead to a higher network capacity. 

\begin{figure}[!t]
\centering
\includegraphics[width=5.5in]{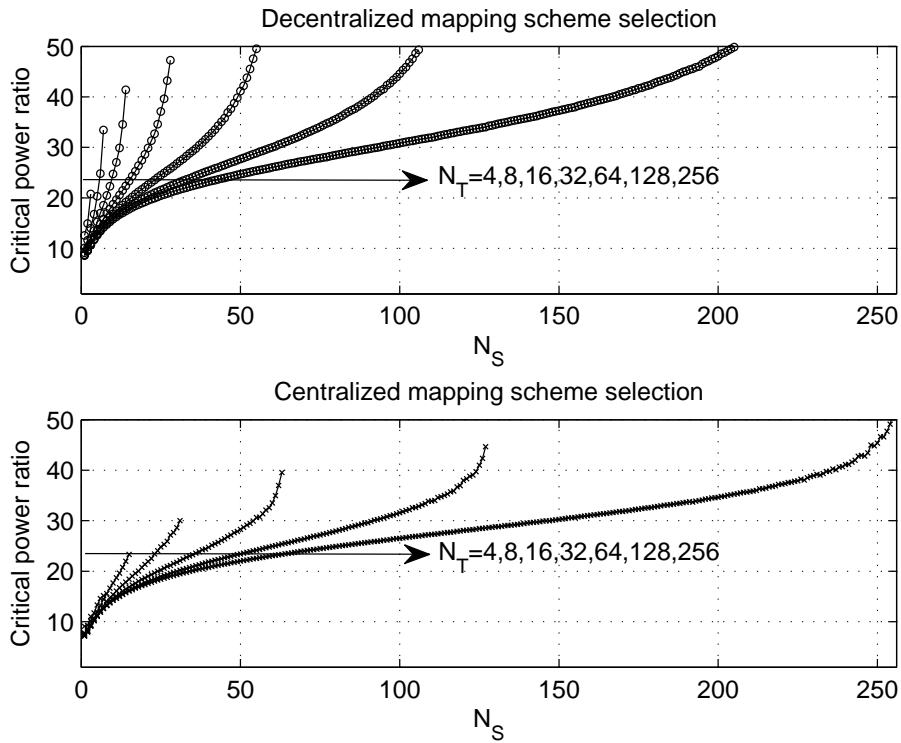}
\caption{Critical power ratio vs. the number of selected subcarriers $N_S$ with different $N_T$ for decentralized and centralized mapping scheme selections.}
\label{fig_critical_power_ratio}
\end{figure}

Besides, both decentralized and centralized adaptive schemes outperform FPSK in terms of average network capacity. However, the proposed adaptive schemes do not always outperform the scheme without adaptation when $P_t/N_0$ is large. This indicates that if the network capacity is regarded as a major system performance metric, our proposed adaptive schemes are more applicable for short-range communication networks, e.g. the underlay D2D networks and the cognitive radio networks for secondary users, in which transmit power is relatively small in order to mitigate the interference to primary users \cite{7809043,7355416}. The ratio of transmit power to noise power by which the capacity without adaptation is equal to that with adaptation (i.e. the crossing point) is termed \textit{critical power ratio}. We have numerically simulated the relation between critical power ratio and $N_S$ with different $N_T$ for both decentralized and centralized cases and have presented the results in Fig. \ref{fig_critical_power_ratio}. From this figure, we can see that the system without adaptation would require a large transmit power in order to have a higher network capacity than that with adaptation, especially when $N_S$ is large. Therefore, this further confirms the applicability of our proposed adaptive schemes in practice.

\subsection{Simulations of error performance}

\begin{figure*}[t!]
    \centering
    \begin{subfigure}[t]{0.5\textwidth}
        \centering
        \includegraphics[width=3.5in]{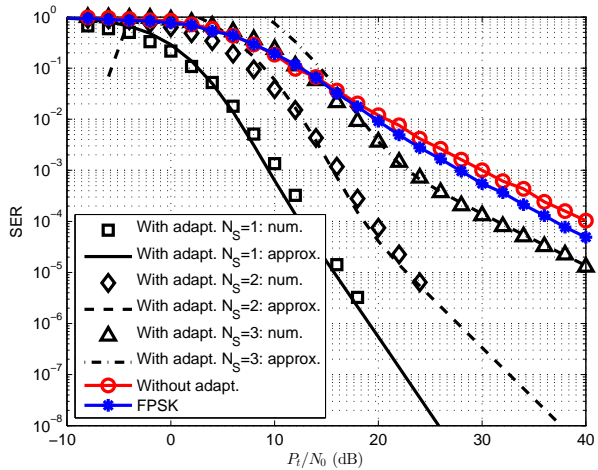}
        \caption{Decentralized mapping scheme selection: $M=2$ (BPSK).}
    \end{subfigure}%
~
    \begin{subfigure}[t]{0.5\textwidth}
        \centering
        \includegraphics[width=3.5in]{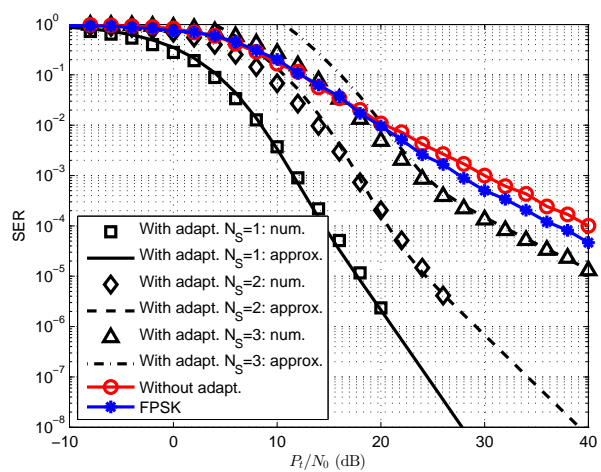}
        \caption{Centralized mapping scheme selection: $M=2$ (BPSK).}
    \end{subfigure}
~
    \centering
    \begin{subfigure}[t]{0.5\textwidth}
        \centering
        \includegraphics[width=3.5in]{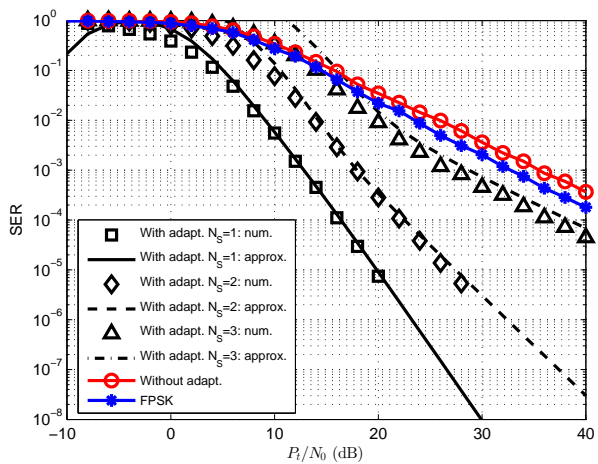}
        \caption{Decentralized mapping scheme selection: $M=4$ (QPSK).}
    \end{subfigure}%
~
    \begin{subfigure}[t]{0.5\textwidth}
        \centering
        \includegraphics[width=3.5in]{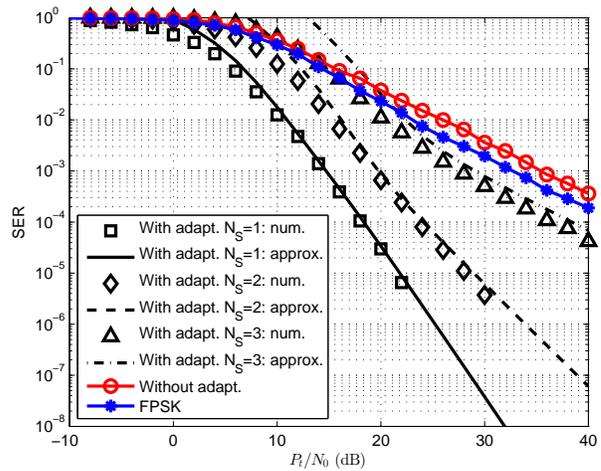}
        \caption{Centralized mapping scheme selection: $M=4$ (QPSK).}
    \end{subfigure}
    \caption{Average SER vs. ratio of transmit power to noise power $P_t/N_0$ for different schemes and $M$.}
    \label{fig_SER_vs_pt}
\end{figure*}

Fixing the total number of subcarrier $N_T=4$, we adopted different numbers of selected subcarriers $N_S$ and APM order $M$, and numerically simulated the relations between average SER and the ratio of transmit power to noise power $P_t/N_0$ for both decentralized and centralized mapping scheme selections. The simulation results are plotted in Fig. \ref{fig_SER_vs_pt}. From this figure, we can see that the derived approximations based on (\ref{dsakdjsakjh2cxqhanshu}) can effectively estimate the SER at high SNR. Also, the diversity orders shown in the simulation results align with our expectation, and are equal to $N_T-N_S$, which is the same as average outage probabilities. In other words, a higher $N_T$ and/or a smaller $N_S$ will lead to a better error performance as expected. Compared with conventional OFDM IM and FPSK schemes, our proposed schemes always exhibit a better error performance when $P_t/N_0$ is relatively large, because of their adaptive mechanisms. Also, OFDM IM with the decentralized mapping scheme selection outperforms OFDM IM with the centralized mapping scheme selection in terms of error performance, as an extra degree of freedom is obtainable by changing the mapping scheme at the relay in the former case. As for the APM order $M$, we can observe a positive correlation between $M$ and $\bar{P}_e$, since the minimum Euclidean distance (MED) is inversely proportional to $M$ when $M$-PSK is adopted for APM.

\section{Conclusion}\label{c}
In this paper, we proposed an adaptive OFDM IM scheme for two-hop DF relay networks, which is capable of obtaining diversity gains by adaptively selecting a portion of the total available OFDM subcarriers, instead of using all for OFDM IM. This provides a possible mechanism allowing reducing throughput for gaining reliability. Considering the complexity and processing capability of relays, we also raised two mapping scheme selection methodologies, which can be applied to different network scenarios. We analyzed a series of important performance metrics of the two-hop networks with OFDM IM, including average outage probability, network capacity and SER. The average outage probability and network capacity can be determined in closed form, and the asymptotic expressions of average outage probability at high SNR were also given in closed form, by which the diversity gain is revealed. Meanwhile, the average SER can be approximated by closed-form expressions, and the approximations are accurate at high SNR. The numerical results substantiated our analysis and also verified the feasibility of the proposed schemes in practice compared to conventional OFDM IM without adaptation as well as FPSK. By the analytical and numerical results provided in this paper, a comprehensive framework for analyzing adaptive OFDM IM in two-hop relay-assisted networks has been constructed, which can be modified to analyze other extended cases with more complicated channel models and transmission protocols.

\appendices
\section{Proof of Lemma \ref{shjdakh2xmznkjca1leemas1}}\label{proofconditionalproba}
When $\mathbf{b}_S(k)$ is transmitted, we need to activate $N_A(k)$ subcarriers from $N_S$ selected subcarriers. Suppose that the worst subcarrier among these $N_A(k)$ active subcarriers in terms of channel gain is the $\xi$th worst in the total ordered $N_T$ subcarriers. Because of the mapping scheme selection, it is easy to derive the relation $N_T-N_S+1\leq\xi\leq N_T-N_A(k)+1$. As a result, by (\ref{dsa411232dsworstous}), the outage probability when the worst subcarrier among these $N_A(k)$ active subcarriers is the $\xi$th worst in the total ordered $N_T$ subcarriers can be written as $F_{i\left(\xi\right)}\left(\frac{sN_0N_A(k)}{P_t}\right)$.

The total number of combinations of $N_A(k)$ active subcarriers is $\binom{N_S}{N_A(k)}$, and the number of possible combinations in which the worst subcarrier among these $N_A(k)$ active subcarriers is the $\xi$th worst is $\binom{N_T-\xi}{N_A(k)-1}$. Consequently, the probability that the worst subcarrier among these $N_A(k)$ active subcarriers is the $\xi$th worst can be determined by
\begin{equation}\small
\Upsilon(k,\xi)=\left.\binom{N_T-\xi}{N_A(k)-1} \middle / \binom{N_S}{N_A(k)} \right.
\end{equation}
Finally, when $\mathbf{b}_S(k)$ is transmitted, the weighted average of the conditional outage probability over $\xi$ gives (\ref{averagewetouts}).

\section{Proof of Lemma \ref{liemadier2}}\label{proofsaikdjakcapa}
Because $\mathbf{\Xi}_1$ and $\mathbf{\Xi}_2$ are independently, discretely and uniformly distributed over domain $\mathcal{D}(k)$, the joint probability mass function (PMF) with respect to $\mathbf{\Xi}_1$ and $\mathbf{\Xi}_2$, i.e. the probability that the $i$th element in $\mathcal{D}(k)$ is chosen for $\mathbf{\Xi}_1$ and the $j$th element in $\mathcal{D}(k)$ is chosen for $\mathbf{\Xi}_2$ is thereby
\begin{equation}\small
p(i,j)=\left(\frac{1}{\mathrm{Card}(\mathcal{D}(k))}\right)^2=\left(\frac{(N_S-N_A(k))!}{N_S!}\right)^2,~\forall~ i,j,
\end{equation}
in which the corresponding network capacity can be written as
\begin{equation}\small
\bar{C}(k|i,j)=\underset{n\in\mathcal{N}_A(k)}{\sum}\Lambda_G\left(\xi_n,\eta_n,\frac{N_0N_A(k)}{P_t}\right).
\end{equation}
Hence, the weighted average of $\bar{C}(k|i,j)$ over $\mathbf{\Xi}_1$ and $\mathbf{\Xi}_2$ is
\begin{equation}\small
\bar{C}(k)=\sum_{\mathbf{\Xi}_1(k)\in\mathcal{D}(k)}\sum_{\mathbf{\Xi}_2(k)\in\mathcal{D}(k)}p(i,j)\bar{C}(k|i,j),
\end{equation}
which yields (\ref{capacituygdsak2s}).

\bibliographystyle{IEEEtran}
\bibliography{bib}

\begin{thebibliography}{10}
\providecommand{\url}[1]{#1}
\csname url@samestyle\endcsname
\providecommand{\newblock}{\relax}
\providecommand{\bibinfo}[2]{#2}
\providecommand{\BIBentrySTDinterwordspacing}{\spaceskip=0pt\relax}
\providecommand{\BIBentryALTinterwordstretchfactor}{4}
\providecommand{\BIBentryALTinterwordspacing}{\spaceskip=\fontdimen2\font plus
\BIBentryALTinterwordstretchfactor\fontdimen3\font minus
  \fontdimen4\font\relax}
\providecommand{\BIBforeignlanguage}[2]{{%
\expandafter\ifx\csname l@#1\endcsname\relax
\typeout{** WARNING: IEEEtran.bst: No hyphenation pattern has been}%
\typeout{** loaded for the language `#1'. Using the pattern for}%
\typeout{** the default language instead.}%
\else
\language=\csname l@#1\endcsname
\fi
#2}}
\providecommand{\BIBdecl}{\relax}
\BIBdecl

\bibitem{proakis}
J.~Proakis and M.~Salehi, \emph{Digital Communications}.\hskip 1em plus 0.5em
  minus 0.4em\relax McGraw-Hill Education, 2007.

\bibitem{6678765}
M.~D. Renzo, H.~Haas, A.~Ghrayeb, S.~Sugiura, and L.~Hanzo, ``Spatial
  modulation for generalized {MIMO}: challenges, opportunities, and
  implementation,'' \emph{Proceedings of the IEEE}, vol. 102, no.~1, pp.
  56--103, Jan. 2014.

\bibitem{6923528}
P.~Banelli, S.~Buzzi, G.~Colavolpe, A.~Modenini, F.~Rusek, and A.~Ugolini,
  ``Modulation formats and waveforms for {5G} networks: Who will be the heir of
  {OFDM}?: An overview of alternative modulation schemes for improved spectral
  efficiency,'' \emph{IEEE Signal Processing Magazine}, vol.~31, no.~6, pp.
  80--93, Nov. 2014.

\bibitem{4382913}
R.~Y. Mesleh, H.~Haas, S.~Sinanovic, C.~W. Ahn, and S.~Yun, ``Spatial
  modulation,'' \emph{IEEE Transactions on Vehicular Technology}, vol.~57,
  no.~4, pp. 2228--2241, Jul. 2008.

\bibitem{6094024}
M.~D. Renzo, H.~Haas, and P.~M. Grant, ``Spatial modulation for
  multiple-antenna wireless systems: a survey,'' \emph{IEEE Communications
  Magazine}, vol.~49, no.~12, pp. 182--191, Dec. 2011.

\bibitem{6587554}
E.~Basar, U.~Aygolu, E.~Panayırcı, and H.~V. Poor, ``Orthogonal frequency
  division multiplexing with index modulation,'' \emph{IEEE Transactions on
  Signal Processing}, vol.~61, no.~22, pp. 5536--5549, Nov. 2013.

\bibitem{7234862}
E.~Basar, ``Multiple-input multiple-output {OFDM} with index modulation,''
  \emph{IEEE Signal Processing Letters}, vol.~22, no.~12, pp. 2259--2263, Dec.
  2015.

\bibitem{6525429}
N.~Serafimovski, A.~Younis, R.~Mesleh, P.~Chambers, M.~D. Renzo, C.~X. Wang,
  P.~M. Grant, M.~A. Beach, and H.~Haas, ``Practical implementation of spatial
  modulation,'' \emph{IEEE Transactions on Vehicular Technology}, vol.~62,
  no.~9, pp. 4511--4523, Nov. 2013.

\bibitem{6823072}
P.~Yang, M.~D. Renzo, Y.~Xiao, S.~Li, and L.~Hanzo, ``Design guidelines for
  spatial modulation,'' \emph{IEEE Communications Surveys Tutorials}, vol.~17,
  no.~1, pp. 6--26, 2015.

\bibitem{7577711}
S.~Dang, G.~Chen, and J.~P. Coon, ``Outage performance analysis of full-duplex
  relay-assisted device-to-device systems in uplink cellular networks,''
  \emph{IEEE Transactions on Vehicular Technology}, vol.~66, no.~5, pp.
  4506--4510, May 2017.

\bibitem{7509396}
E.~Basar, ``Index modulation techniques for {5G} wireless networks,''
  \emph{IEEE Communications Magazine}, vol.~54, no.~7, pp. 168--175, Jul. 2016.

\bibitem{7414384}
M.~Agiwal, A.~Roy, and N.~Saxena, ``Next generation {5G} wireless networks: A
  comprehensive survey,'' \emph{IEEE Communications Surveys Tutorials},
  vol.~18, no.~3, pp. 1617--1655, 2016.

\bibitem{6473920}
B.~Han, M.~Peng, Z.~Zhao, and W.~Wang, ``A multidimensional resource-allocation
  optimization algorithm for the network-coding-based multiple-access relay
  channels in {OFDM} systems,'' \emph{IEEE Transactions on Vehicular
  Technology}, vol.~62, no.~8, pp. 4069--4078, Oct. 2013.

\bibitem{4595667}
Z.~Han, T.~Himsoon, W.~P. Siriwongpairat, and K.~J.~R. Liu, ``Resource
  allocation for multiuser cooperative {OFDM} networks: Who helps whom and how
  to cooperate,'' \emph{IEEE Transactions on Vehicular Technology}, vol.~58,
  no.~5, pp. 2378--2391, Jun. 2009.

\bibitem{7445895}
S.~Dang, J.~P. Coon, and G.~Chen, ``An equivalence principle for {OFDM}-based
  combined bulk/per-subcarrier relay selection over equally spatially
  correlated channels,'' \emph{IEEE Transactions on Vehicular Technology},
  vol.~66, no.~1, pp. 122--133, Jan. 2017.

\bibitem{5733966}
A.~Chandra, C.~Bose, and M.~K. Bose, ``Wireless relays for next generation
  broadband networks,'' \emph{IEEE Potentials}, vol.~30, no.~2, pp. 39--43,
  Mar. 2011.

\bibitem{7120187}
S.~Narayanan, M.~D. Renzo, F.~Graziosi, and H.~Haas, ``Distributed spatial
  modulation: A cooperative diversity protocol for half-duplex relay-aided
  wireless networks,'' \emph{IEEE Transactions on Vehicular Technology},
  vol.~65, no.~5, pp. 2947--2964, May 2016.

\bibitem{5956586}
N.~Serafimovski, S.~Sinanovic, M.~D. Renzo, and H.~Haas, ``Dual-hop spatial
  modulation ({Dh-SM}),'' in \emph{Proc. IEEE VTC Spring}, Budapest, Hungary,
  May 2011.

\bibitem{5752793}
P.~Yang, Y.~Xiao, Y.~Yu, and S.~Li, ``Adaptive spatial modulation for wireless
  {MIMO} transmission systems,'' \emph{IEEE Communications Letters}, vol.~15,
  no.~6, pp. 602--604, Jun. 2011.

\bibitem{7055353}
X.~Wu, M.~D. Renzo, and H.~Haas, ``Adaptive selection of antennas for optimum
  transmission in spatial modulation,'' \emph{IEEE Transactions on Wireless
  Communications}, vol.~14, no.~7, pp. 3630--3641, Jul. 2015.

\bibitem{6423761}
R.~Rajashekar, K.~V.~S. Hari, and L.~Hanzo, ``Antenna selection in spatial
  modulation systems,'' \emph{IEEE Communications Letters}, vol.~17, no.~3, pp.
  521--524, Mar. 2013.

\bibitem{7763523}
Y.~Wang, W.~Liu, M.~Jin, S.~Jang, and J.~M. Kim, ``{FQAM/FPSK} modulation for
  spatial modulation systems,'' in \emph{Proc. IEEE ICTC}, Jeju, Korea, Oct.
  2016.

\bibitem{6162549}
D.~Tsonev, S.~Sinanovic, and H.~Haas, ``Enhanced subcarrier index modulation
  {(SIM) OFDM},'' in \emph{Proc. IEEE GLOBECOM}, Houston, Texas, USA, Dec.
  2011.

\bibitem{tse2005fundamentals}
D.~Tse and P.~Viswanath, \emph{Fundamentals of Wireless Communication}, ser.
  Wiley series in telecommunications.\hskip 1em plus 0.5em minus 0.4em\relax
  Cambridge University Press, 2005.

\bibitem{6142142}
M.~D. Renzo and H.~Haas, ``Bit error probability of {SM-MIMO} over generalized
  fading channels,'' \emph{IEEE Transactions on Vehicular Technology}, vol.~61,
  no.~3, pp. 1124--1144, Mar. 2012.

\bibitem{7330022}
M.~Wen, X.~Cheng, M.~Ma, B.~Jiao, and H.~V. Poor, ``On the achievable rate of
  {OFDM} with index modulation,'' \emph{IEEE Transactions on Signal
  Processing}, vol.~64, no.~8, pp. 1919--1932, Apr. 2016.

\bibitem{luna2013constellation}
J.~Luna-Rivera, D.~U. Campos-Delgado, and M.~Gonzalez-Perez, ``Constellation
  design for spatial modulation,'' \emph{Procedia Technology}, vol.~7, pp.
  71--78, 2013.

\bibitem{gonzalez2012pre}
M.~Gonz{\'a}lez-P{\'e}rez, J.~Luna-Rivera, and D.~Campos-Delgado,
  ``Pre-equalization for {MIMO} wireless systems using spatial modulation,''
  \emph{Procedia Technology}, vol.~3, pp. 1--8, 2012.

\bibitem{7247508}
S.~Dang, J.~P. Coon, and D.~E. Simmons, ``Combined bulk and per-tone relay
  selection in super dense wireless networks,'' in \emph{Proc. IEEE ICC},
  London, UK, Jun. 2015.

\bibitem{7469311}
N.~Ishikawa, S.~Sugiura, and L.~Hanzo, ``Subcarrier-index modulation aided
  {OFDM} - will it work?'' \emph{IEEE Access}, vol.~4, pp. 2580--2593, 2016.

\bibitem{6211487}
X.~Tao, X.~Xu, and Q.~Cui, ``An overview of cooperative communications,''
  \emph{IEEE Communications Magazine}, vol.~50, no.~6, pp. 65--71, Jun. 2012.

\bibitem{6157252}
W.~Yang and Y.~Cai, ``On the performance of the block-based selective {OFDM}
  decode-and-forward relaying scheme for {4G} mobile communication systems,''
  \emph{Journal of Communications and Networks}, vol.~13, no.~1, pp. 56--62,
  Feb. 2011.

\bibitem{5071270}
W.~Wang and R.~Wu, ``Capacity maximization for {OFDM} two-hop relay system with
  separate power constraints,'' \emph{IEEE Transactions on Vehicular
  Technology}, vol.~58, no.~9, pp. 4943--4954, Nov. 2009.

\bibitem{4601434}
J.~Jeganathan, A.~Ghrayeb, and L.~Szczecinski, ``Spatial modulation: optimal
  detection and performance analysis,'' \emph{IEEE Communications Letters},
  vol.~12, no.~8, pp. 545--547, Aug. 2008.

\bibitem{6678968}
G.~Chen, O.~Alnatouh, and J.~Chambers, ``Outage probability analysis for a
  cognitive amplify-and-forward relay network with single and multi-relay
  selection,'' \emph{IET Communications}, vol.~7, no.~17, pp. 1974--1981, Nov.
  2013.

\bibitem{789668}
M.~S. Alouini and A.~J. Goldsmith, ``A unified approach for calculating error
  rates of linearly modulated signals over generalized fading channels,''
  \emph{IEEE Transactions on Communications}, vol.~47, no.~9, pp. 1324--1334,
  Sept. 1999.

\bibitem{1188428}
M.~Chiani and D.~Dardari, ``Improved exponential bounds and approximation for
  the {Q}-function with application to average error probability computation,''
  in \emph{Proc. IEEE GLOBECOM}, Nov. 2002.

\bibitem{1197843}
L.~Zheng and D.~N.~C. Tse, ``Diversity and multiplexing: a fundamental tradeoff
  in multiple-antenna channels,'' \emph{IEEE Transactions on Information
  Theory}, vol.~49, no.~5, pp. 1073--1096, May 2003.

\bibitem{7809043}
S.~Dang, J.~P. Coon, and G.~Chen, ``Resource allocation for full-duplex
  relay-assisted device-to-device multicarrier systems,'' \emph{IEEE Wireless
  Communications Letters}, vol.~6, no.~2, pp. 166--169, Apr. 2017.

\bibitem{7355416}
G.~Chen, Y.~Gong, P.~Xiao, and J.~A. Chambers, ``Dual antenna selection in
  secure cognitive radio networks,'' \emph{IEEE Transactions on Vehicular
  Technology}, vol.~65, no.~10, pp. 7993--8002, Oct. 2016.

\end{thebibliography}

\end{document}